\documentclass[twocolumn,tighten]{aastex62}

\hypersetup{colorlinks,%
 citecolor=blue,%
 linkcolor=blue}
\definecolor{lightblue}{rgb}{1,.5,0}

\newcommand{\ror}\textbf{}

\received{\today}
\revised{\today}
\accepted{\today}
\submitjournal{ApJS}

\shorttitle{Dwarf-Dwarf mergers}
\shortauthors{Paudel et al.}

\begin{document}
\title{A CATALOG OF MERGING DWARF GALAXIES IN THE LOCAL UNIVERSE}

\author[0000-0003-2922-6866]{Sanjaya Paudel}
\affil{Department of Astronomy and Center for Galaxy Evolution Research, Yonsei University, Seoul 03722, Korea}
\email{sanjpaudel@gmail.com (SP)\\
rorysmith@kasi.re.kr (RS)\\
sjyoon0691@yonsei.ac.kr (SJY)\\
ashniet.caskortish@gmail.com (PCC)\\
pierre-alain.duc@astro.unistra.fr (PAD)}
\author{Rory Smith}
\affiliation{Korea Astronomy and Space Science Institute, Daejeon 305-348, Republic of Korea}
\author{ Suk Jin Yoon}
\affiliation{Department of Astronomy and Center for Galaxy Evolution Research, Yonsei University, Seoul 03722, Korea}
\author{Paula Calder\'on-Castillo}
\affiliation{Astronomy Department, Universidad de Concepci\'on, Casilla 160-C, Concepci\'on, Chile}
\author{Pierre-Alain Duc}
\affiliation{Universite\' de Strasbourg, CNRS, Observatoire astronomique de Strasbourg, UMR 7550, F-67000 Strasbourg, France}

\begin{abstract}

We present the largest publicly available catalog of interacting dwarf galaxies. It includes 177 nearby merging dwarf galaxies of stellar mass M$_{*}$ $<$ 10$^{10}$M$_{\sun}$ and redshifts z $<$ 0.02. These galaxies are selected by visual inspection of publicly available archival imaging from two wide-field optical surveys (SDSS III and the Legacy Survey), and they possess low surface brightness features that are likely the result of an interaction between dwarf galaxies. We list UV and optical photometric data which we use to estimate stellar masses and star formation rates. So far, the study of interacting dwarf galaxies has largely been done on an individual basis, and lacks a sufficiently large catalog to give statistics on the properties of interacting dwarf galaxies, and their role in the evolution of low mass galaxies. We expect that this public catalog can be used as a reference sample to investigate the effects of the tidal interaction on the evolution of star-formation, morphology/structure of dwarf galaxies. 

Our sample is overwhelmingly dominated by star-forming galaxies, and they are generally found significantly below the red-sequence in the color-magnitude relation. The number of early-type galaxies is only 3 out of 177. We classify them, according to observed low surface brightness features, into various categories including shells, stellar streams, loops, antennae or simply interacting. We find that dwarf-dwarf interactions tend to prefer the low density environment. Only 41 out of the 177 candidate dwarf-dwarf interaction systems have giant neighbors within a sky projected distance of 700 kpc and a line of sight radial velocity range $\pm$700 km/s and, compared to the LMC-SMC, they are generally located at much larger sky-projected distances from their nearest giant neighbor.

\end{abstract}


\section{Introduction}\label{intro}
A plethora of observational studies now support the conclusion that mergers between galaxies are frequent phenomena. In the $\Lambda$CDM cosmology \citep{Spergel07}, the assembly of large scale structure happens in a hierarchical fashion, and mergers play a fundamental role in both the growth and evolution of galaxies \citep{Conselice09}. Both observations and numerical simulations concur that massive elliptical galaxies were likely formed predominantly by the mergers of disk galaxies \citep{Springel05,Naab07,Duc11,Duc15}.

On the other hand, it is a common belief that the shallow potential well of  low mass galaxies causes them to be more sensitive to their surrounding environment than massive galaxies. Dwarf galaxies exhibit a strong morphological segregation: the most evolved / oldest dwarf galaxies (i.e dwarf Spheroidal (dSph) or dwarf early-type (dE)) are found exclusively in the group and cluster environments \citep{Kormendy09,Lisker09,Boselli06}. Meanwhile  dwarfs with on-going  star-formation activity (such as Blue Compact Dwarf galaxies (BCDs, \citealt{Gil03,Papaderos96}) or dwarf irregulars (dIrs, \citealt{Gallagher84}) are mainly found in less dense environments. Indeed, a study of the environmental dependence on the star-formation activity in dwarf galaxies by \cite{Geha12} concluded that early-type dwarf galaxies (10$^{6}$ $<$ M$_{*}$  $<$ 10$^{9}$ ) are extremely rare in the field. The origin of the different dwarf galaxy types and the possible evolutionary links between them are the subject of much research and debate \citep{Lisker09}.  

The evolution of dwarf galaxies throughout the merging process has yet to be explored in detail. However, in the last few years the observational evidence for mergers between dwarf galaxies has been growing \citep[e.g.][]{Amorisco14,Crnojevic14,Delgado12,Johnson13,Nidever13,Rich12,Paudel17}. The possibilities that certain low mass early-type galaxy (or dEs) might also be formed through mergers, similar to massive ellipticals, has been speculated in order to explain peculiar observational properties such as kinematically decoupled cores and boxy shape isophotes \citep{Toloba14,Graham12,Geha05}. If this is the case, one might expect the progenitors of some dEs to exhibit characteristic features that arise during mergers, such as tidal debris.

Much work has been done to understand the physical processes driving galaxy evolution in the merger of massive galaxies. It has been shown by many observational and theoretical studies that, during the intermediate phases of interactions, large scale tidal interactions trigger the formation of peculiar features like shells, streams, bridges and tails \citep{Toomre72,Eneev73,Barnes09,Struck12,Duc13}. The presence of such structures,which is also predicted by numerical simulations, is now frequently observed in deep imaging surveys \citep{Conselice99,Smith07,Tal09,Duc11,Duc14b,Kim12,Struck99,vanDokkum05}.

In the low mass regime, a detailed study of interacting systems has been exceptionally rare. This is likely because such systems are not as easy to observe as in massive systems. A part of the reason for this could perhaps be that the tidal features which are produced are not as spectacular as in merging giant galaxies due to the relatively weak tidal forces acting upon them. But certainly dwarf galaxies, by nature, are inherently low surface brightness systems and thus the tidal feature emerging from them are often even more low-surface brightness, making them challenging to detect. Only recently, with the advent of low surface brightness imaging techniques, and dedicated data reduction procedures, have we been able to better detect such features \citep{Abraham14,Duc14,Mihos17}.

 Dwarf-dwarf interactions might also be distinct from giant-giant interactions for another reason. In low density environments, dwarfs are often much more gas rich than giant galaxies. Furthermore, the dynamics of gas is not scalable in the same way that the dissipationless star and dark matter components are. For example, the neutral hydrogen in galaxies has a typical velocity dispersion of $\sim$10~km/s. For giants, with rotation velocities of more than 100 km/s, this internal velocity may have a minor contribution to the overall disk dynamics. However for dwarf galaxies, a 10 km/s velocity dispersion can make a significant contribution to the internal dynamics. This may potentially lead to a difference in the star-formation efficiency and overall evolutionary history of dwarf galaxies compared to giants.

A few detailed observational studies of some individual dwarf galaxies with merging feature have been reported in recent years \citep{Rich12,Paudel15,Pearson16,Annibali16}. In our nearby vicinity, apart from the infamous interaction between the Magellanic clouds, there is also NGC 4449, an ongoing interaction between a Magellanic type dwarf and it's nearby dwarf companions \citep{Putman03,Delgado12,Rich12,Besla16} in which a small stretched stellar stream is observed at the edge of the NGC 4449. The presence of a shell feature in the Fornax dwarf spheroidal has also been interpreted as a relic of a recent merger \citep{Coleman04,Yozin12}. In addition to this, \cite{Paudel15} reported interactions between dwarf galaxies where the overall morphological appearance is similar to that of the well known giant system Arp 104. Also there is UM 448, a merging blue compact dwarf galaxy (BCD), which possesses a pronounced tidal tail that was studied in \cite{James13}.  

 \begin{figure*}
\label{samclas}
\includegraphics[width = 17cm]{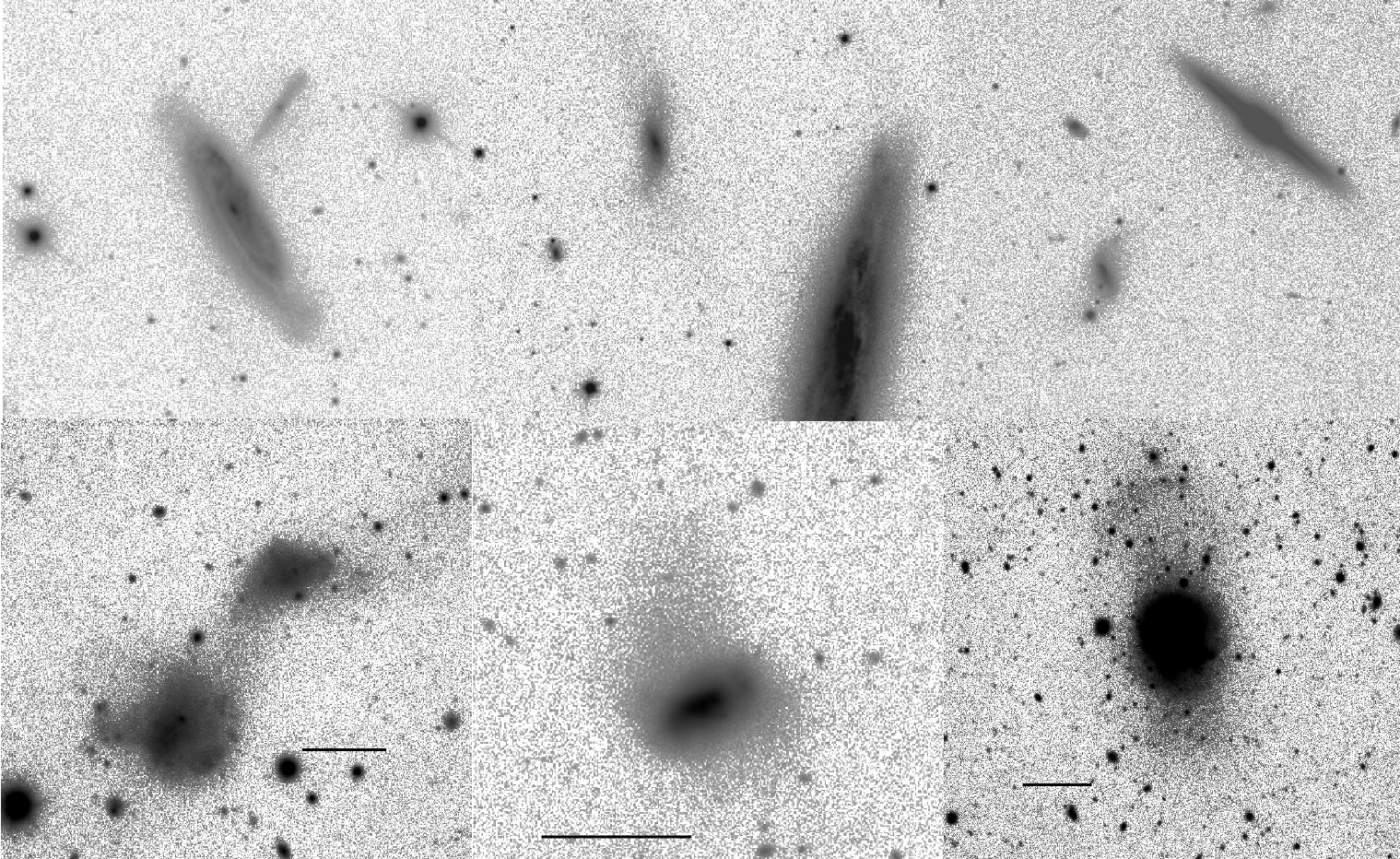}
\caption{Representative examples of dwarf galaxies with tidal features. \\
Top: Examples where we conclude a dwarf galaxy is interacting with, and being deformed by the tidal field of a nearby giant galaxy and they have been excluded from the catalog.
\\ Bottom: Examples of dwarfs that we classify as having interacted with another dwarf, categorized into three different types of tidal features (i.e. from left to right; interacting, tidal tail and shell features). For all images, the black horizontal bar represents a scale of 30\arcsec.}
\end{figure*}

Despite these detailed studies of a few intriguing examples, very little is known about whether these systems are representative of dwarf-dwarf interactions in general. Nevertheless, given that the majority of galaxies in the Universe are dwarfs, it is clearly important to know how dwarf galaxies evolve through the merging process. Dwarf galaxies not only differ in mass from giant galaxies, but they also have higher gas mass fractions and lower star-formation efficiencies. Low mass galaxies are also typically dominated by exponential disks. How might these properties affect the interaction compared to their giant counterparts? Despite the very similar visual morphology of UGC 6741 system to Arp 104, \cite{Paudel15} reported a number of star forming region in the bridge connecting the two interacting galaxies, whereas such star formation is completely absent in Arp104 \citep{Gallagher10}. Dwarf galaxies, by definition can exert lower tidal forces compared to their massive counterparts -- does this result in differences in the tidal features compared to those produced by the much stronger tidal forces of giant galaxies? Antenne (NGC 4038/39), Mice (NGC 4676), Tadpole (UGC 10214) and Guitar (NGC 5291) are some spectacular examples of tidal features that we observe in the interactions between giant galaxies. In addition to this, prominent shell features (e.g. NGC 747 or NGC 7600) are also commonly observed in giant elliptical galaxies \citep{Duc15}.

\begin{figure*}
\includegraphics[width = 18cm]{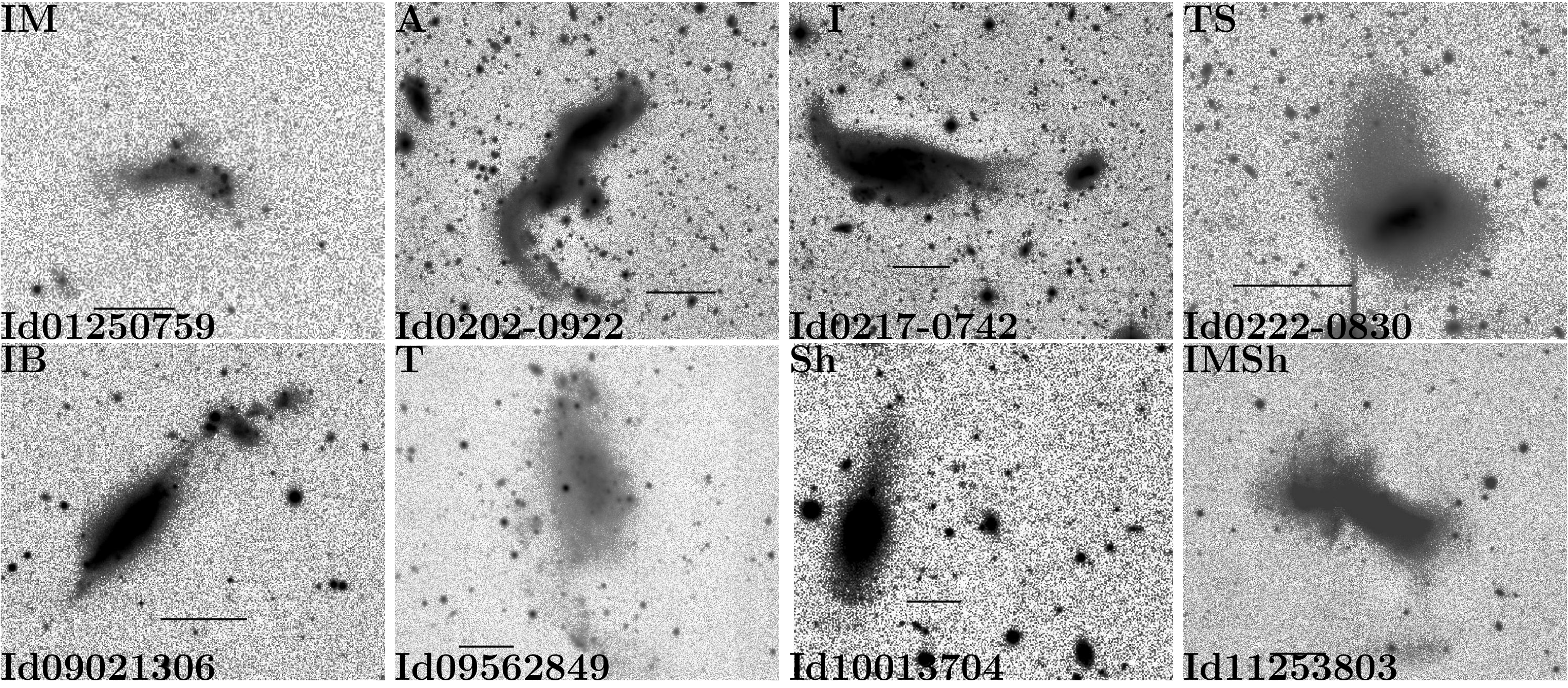}
\caption{Representative examples of the different morphological classes by which we categorize our merging dwarf galaxies. The field of view and color stretching is arbitrarily chosen to make best view of both interacting galaxies and low-surface brightness features.  An image scale of 30" is shown by the black horizontal bar. See \S\ref{class}  for further details. The complete list of images is shown in Figure \ref{fcat}.}
\label{mclass}
\end{figure*}

Recently, a systematic study of dwarf galaxy pairs, likely to be interacting, in the SDSS data base has been presented by \cite{Stierwalt15}, although the full catalog of 104 dwarf-dwarf pair galaxies with the name and position of the galaxies has yet to be publicly released. They are mostly gas-rich and star forming systems, located in low density environments. A sub-set of this sample is studied in \cite{Pearson16}, where their HI morphology is analyzed. They find an extended HI morphology in their tidally interacting galaxy sample compared to non-paired analogues. In this work, we focus on the optical morphology of dwarf-dwarf galaxy interactions. For this, we first create a sample of interacting dwarf galaxies based on a visual analysis of color images from the Sloan Digital Sky Survey (SDSS). We have  conducted a systematic search for dwarf galaxies possessing tidal feature, such as stellar streams, shells or filaments, through a careful examination of the SDSS images. Although these features could also be produced by interactions with other galaxies, in this work we try to focus on a sample of dwarf galaxies with tidal features that are likely produced by dwarf-dwarf mergers.

Without aiming to provide detailed science of dwarf-dwarf mergers in this study, we instead aim to provide a sample of dwarf-dwarf merging systems that can later be used for more detailed science. Given the fairly good number statistics of our sample, we also attempt to understand their typical environment.

For this work, we adopt a standard cosmological model with the following parameters: H$_{0}$ = 71 kms$^{-1}$ Mpc$^{-1}$,  $\Omega$$_{m}$ = 0.3 and $\Omega$$_{\Lambda}$= 0.7.

\section{Sample Selection}

\subsection{Selection of dwarf-dwarf interactions}\label{selection}
Our main aim is to create a large catalog of merging dwarf galaxies. We are mostly interested in dwarf galaxies with tidal features that are likely to be produced by interacting/merging dwarf galaxies. We first search for such disrupted candidates in the large imaging database of the SDSS and the Legacy survey\footnote{http://legacysurvey.org}.

For this, we use a visual inspection of the true color images from the SDSS-III \citep{Aihara11} and the Legacy survey \citep{Blum16}. The parent sample of galaxies is drawn using a query in the NED where we select galaxies within a redshift range of z $<$ 0.02 from the region of sky covered by the SDSS and Legacy survey. We start by selecting galaxies of magnitude M$_{r}$ $>$-19 mag to ensure the parent sample of galaxies is predominantly composed of dwarf galaxies. However, note that this magnitude cutoff is only to select the parent sample and we apply a further stellar mass constrain to select the final sample. The stellar mass of candidate galaxies in this sample are measured from our own photometric measurement as described in \S \ref{data}. The total number of galaxies in this redshift range is $\approx$20,000.

We then extract a cut-out color image from the SDSS sky-server and Legacy survey. As our prime goal is to find tidal debris around the dwarf galaxies, we first collect a sample of dwarf galaxies with observed tidal debris, without considering the origin of the debris at this stage. As might be expected, the majority of the tidal feature are created by interactions with their neighbor giant galaxies. This large sample of disrupted galaxies or galaxies that exhibit tidal debris contain more than 700 candidates. However, for this particular work, we focus on dwarf-dwarf interactions. Another comprehensive catalog of tidally interacting dwarf galaxies with nearby giant galaxies, similar to those studied in \cite{Paudel14}, will be published later (Paudel et al in preparation).

Our visual inspection process involves multiple steps. First, we look for any signature of tidal features in the true color images. If a hint is found, we then re-examine the coadded fits file of the multiple bands available in the archive. The coaddition provides higher SNR than the single band images. Additionally, we also search for the availability of deeper images in various publicly available archives. In this regard, the archival images of the CHFT\footnote{http://www.cfht.hawaii.edu/} were very helpful for visual confirmation of the presence of low surface brightness feature around dwarf galaxies. From the CHFT archive, we use the Megapipe stack\footnote{http://www.cadc-ccda.hia-iha.nrc-cnrc.gc.ca/en/megapipe/} produced by The Elixir System \citep{Gwyn08}. Megapipe stack images are a pipeline reduced images of CHFT MegaCam observation. 

 Finally, we classify the dwarf galaxies with tidal feature into two broad categories; dwarf-dwarf interaction/merger  and dwarf-giant interactions. We show examples of these two classes in Figure \ref{samclas}. The first row shows images of the tidal distortion of dwarf galaxies by nearby giant galaxies and in the second row we show examples of merging dwarf galaxies. It is not always trivial to determine if the observed tidal features were created by merging dwarf galaxies except when the interacting pair have not completely merged yet -- like, for example, in the Antennae-like dwarf galaxies (see lower left panel of Figure \ref{samclas}) or simply interacting pairs (see lower middle panel of Figure \ref{samclas}).  However, from our past experience, we can often suspect a particular origin according to the appearance of the observed low-surface brightness tidal features. For example, we have shown that shell features about dwarf galaxies are well produced by a merger origin \citep{Paudel17}. Meanwhile, an S-shaped elongated stellar envelope is likely to be produced by tidal stretching from a nearby giant galaxy \citep{Paudel13,Paudel14b}. These selection criteria are indeed subjective. But, we are keen to avoid including dwarfs that are interacting with a giant galaxy in this catalog. This may create a bias against merged dwarfs near giants, see discussion \S\ref{cons}.

 \subsection{Sample classification}\label{class}
 The final sample consists of 177 systems with a limit in the combined stellar mass of the system of $<$10$^{10}$ M$_{\sun}$. We further classify these objects according to the morphology of their tidal features, mainly grouping them into three categories; Interacting, Shell and Tidal tail features. In addition to this, in some cases, we further sub-classify them according to the details of observed low surface brightness feature, see below.

\startlongtable
\begin{deluxetable*}{c|ccRccccccc}
\tablecaption{Physical properties merging dwarf galaxies. \label{mtb}}
\tablenum{1}
\tablecolumns{11}
\tablehead{
\colhead{No.} & 
\colhead{ID} & 
\colhead{RA} & 
\colhead{Dec} & 
\colhead{z} & 
\colhead{m$_{g}$} &
\colhead{m$_{r}$} &
\colhead{m$_{FUV}$} &
\colhead{m$_{NUV}$} &
\colhead{Feature} &
\colhead{Galaxy}\\
\colhead{} &
\colhead{} &
\colhead{deg} &
\colhead{deg} &
\colhead{} &
\colhead{mag} &
\colhead{mag} &
\colhead{mag} &
\colhead{mag} &
\colhead{} &
\colhead{name} 
}
\startdata
001 & Id01130052 & 018.4138 & 00.8741 & 0.0039 & 15.87 & 15.76 & 16.79 & 16.96    & 	I 	      &  UGC 00772  \\
002 & Id01250759 & 021.3957 & 07.9908 & 0.0097 & 15.18 & 15.14 & 15.89 & 15.89    & 	IM	      &  UGC 00993  \\
003 & Id01482838 & 027.1545 & 28.6427 & 0.0125 & 15.78 & 15.35 & 17.28 & 16.64    & 	IB	      &     \\
004 & Id0155-0011 & 028.9989 & -0.1855 & 0.0121 & 16.47 & 16.01 & 17.76 & 16.55    & 	Sh	      &     \\
005 & Id0202-0922 & 030.6615 & -9.3703 & 0.0180 & 15.43 & 15.13 & 17.38 & 17.18    & 	A	      &  PGC007782   \\
006 & Id02032202 & 030.8279 & 22.0441 & 0.0088 & 13.86 & 13.55 & 14.15 & 14.64    & 	IM	      &  UGC 01547  \\
007 & Id0210-0124 & 032.5408 & -1.4013 & 0.0119 & 15.03 & 14.83 & 15.63 & 16.18    & 	E	      &  KUG 0207-016A \\
008 & Id0217-0742 & 034.3948 & -7.7040 & 0.0160 & 14.76 & 14.45 & 15.68 & 16.18    & 	I	      &  PGC008757   \\
009 & Id0221-0928 & 035.4799 & -9.4766 & 0.0123 & 16.01 & 15.81 & 17.20 & 16.64    & 	I	      &     \\
010 & Id0222-0830 & 035.5498 & -8.5101 & 0.0156 & 15.25 & 14.90 & 17.15 & 17.04    & 	TS	      &     \\
011 & Id0227-0837 & 036.9460 & -8.6261 & 0.0167 & 16.41 & 16.09 & 17.03 & 17.70    & 	TS	      &     \\
012 & Id02430338 & 40.82292 & 3.64472 & 0.0140 & 15.08 & 14.94 & \nodata & \nodata          &   I        &   PGC010297\\
013 & Id07183123 & 109.6395 & 31.3866 & 0.0114 & 14.41 & 14.02 & \nodata & \nodata          &   I        &  \\
014 & Id07551505 & 118.8437 & 15.0938 & 0.0154 & 14.77 & 14.50 & 16.37 & 15.71    & 	IM	      &  PGC022184   \\
015 & Id08012517 & 120.3283 & 25.2899 & 0.0155 & 15.40 & 15.18 & \nodata & \nodata          & 	I	      &  PGC022495   \\
016 & Id08092137 & 122.4474 & 21.6215 & 0.0111 & 15.98 & 15.75 & 16.42 & 15.34    & 	T	      &     \\
017 & Id08114627 & 122.7846 & 46.4656 & 0.0074 & 13.87 & 13.69 & \nodata & 14.63       & 	I	      &  PGC022955   \\
018 & Id08213419 & 125.4696 & 34.3272 & 0.0077 & 16.49 & 16.31 & \nodata & \nodata          & 	I	      &     \\
019 & Id08291427 & 127.3861 & 14.4518 & 0.0197 & 16.02 & 15.72 & 16.86 & 16.42    & 	I	      &     \\
020 & Id08331920 & 128.3229 & 19.3466 & 0.0193 & 14.97 & 14.68 & 16.54 & 16.00    & 	IM	      &     \\
021 & Id08332932 & 128.3457 & 29.5386 & 0.0069 & 13.06 & 12.73 & 14.64 & 14.35    & 	I	      &     \\
022 & Id08350340 & 128.8927 & 03.6717 & 0.0131 & 16.96 & 16.71 & \nodata & \nodata          & 	T	      &     \\
023 & Id08360509 & 129.1278 & 05.1659 & 0.0135 & 15.56 & 15.40 & 16.68 & 15.94    & 	I	      &     \\
024 & Id0851-0221 & 132.9080 & -2.3660 & 0.0109 & 13.62 & 13.38 & 14.86 & 15.28    & 	I	      &  UGC 04638  \\
025 & Id08580619 & 134.6239 & 06.3213 & 0.0119 & 15.51 & 15.14 & 16.70 & 16.23    & 	IB	      &  UGC 04703  \\
026 & Id09003543 & 135.0654 & 35.7276 & 0.0101 & 13.47 & 13.26 & 15.04 & 14.71    & 	I	      &  NGC 2719  \\
027 & Id09002536 & 135.0999 & 25.6147 & 0.0060 & 14.24 & 14.14 & 15.82 & 15.49    & 	T	      &     \\
028 & Id09021306 & 135.6726 & 13.1077 & 0.0164 & 14.78 & 14.48 & 16.55 & 16.00    & 	IB	      &  PGC025403   \\
029 & Id09114239 & 137.7848 & 42.6562 & 0.0060 & 15.56 & 15.31 & 16.39 & 15.37    & 	I	      &     \\
030 & Id09164259 & 139.1047 & 42.9916 & 0.0085 & 14.77 & 14.57 & \nodata & \nodata          & 	I	      &  PGC026162   \\
031 & Id09165946 & 139.1834 & 59.7746 & 0.0137 & 14.88 & 14.60 & 16.10 & 15.93    & 	I	      &  MRK 0019  \\
032 & Id09201920 & 140.1685 & 19.3374 & 0.0139 & 15.96 & 15.70 & 17.09 & 16.58    & 	I	      &     \\
033 & Id09296627 & 142.2739 & 66.4579 & 0.0115 & 14.97 & 14.75 & 16.37 & 16.03    & 	I	      &  UGC 05042  \\
034 & Id09306026 & 142.5268 & 60.4481 & 0.0136 & 15.21 & 15.07 & 16.02 & 15.74    & 	I	      &     \\
035 & Id09333336 & 143.4291 & 33.6002 & 0.0052 & 15.32 & 14.98 & 16.77 & 16.83    & 	Sh	      &  KUG 0930+338  \\
036 & Id09381942 & 144.5608 & 19.7111 & 0.0144 & 17.19 & 16.99 & \nodata & \nodata          & 	Sh	      &     \\
037 & Id09420929 & 145.7212 & 9.49164 & 0.0107 & 14.26 & 14.13 & 15.31 & 15.07    & 	I	      &  UGC 05189  \\
038 & Id0944-0039 & 146.0300 & -0.6598 & 0.0041 & 14.85 & 14.65 & 16.09 & 15.62    & 	I	      &  UGC 05205  \\
039 & Id09494402 & 147.2779 & 44.0477 & 0.0156 & 15.88 & 15.76 & 16.68 & 15.97    &    I        &  \\
040 & Id09514419 & 147.9137 & 44.3190 & 0.0150 & 16.18 & 15.99 & 17.03 & 16.79    & 	I	      &     \\
041 & Id09516853 & 147.9874 & 68.8841 & 0.0146 & 16.93 & 16.49 & 19.10 & 19.40    & 	I	      &      \\
042 & Id09530702 & 148.4811 & 07.0465 & 0.0174 & 16.23 & 15.98 & 17.79 & 17.85    & 	L	      &     \\
043 & Id09550823 & 148.8737 & 8.39062 & 0.0041 & 14.77 & 14.38 & 16.33 & 15.49    & 	Sh	      &  UGCA 188  \\
044 & Id09562849 & 149.1918 & 28.8288 & 0.0015 & 14.38 & 14.33 & 15.56 & 15.49    & 	T	      &     \\
045 & Id10004531 & 150.0195 & 45.5198 & 0.0056 & 16.50 & 16.35 & \nodata & \nodata          & 	I	      &  KUG 0956+457  \\
046 & Id10004311 & 150.0242 & 43.1919 & 0.0056 & 15.31 & 15.02 & \nodata & \nodata          & 	Sh	      &     \\
047 & Id10013704 & 150.3099 & 37.0709 & 0.0048 & 15.33 & 15.19 & 16.71 & 16.10    & 	Sh	      &  PGC029004   \\
048 & Id1007-0631 & 151.8945 & -6.5232 & 0.0158 & 15.31 & 14.99 & 16.30 & 16.76    & 	IM	      &     \\
049 & Id10080227 & 152.0430 & 02.4634 & 0.0068 & 14.96 & 14.24 & 17.93 & 16.82    & 	E	      &  PGC029471   \\
050 & Id10100509 & 152.6575 & 05.1502 & 0.0137 & 14.85 & 14.48 & 16.38 & 17.18    & 	S	      &  CGCG 036-048  \\
051 & Id10170419 & 154.2989 & 04.3312 & 0.0045 & 15.63 & 15.35 & 17.40 & 17.22    & 	IB	      &  UGC 05551  \\
052 & Id10174308 & 154.3874 & 43.1448 & 0.0037 & 16.14 & 15.80 & \nodata & \nodata          & 	Sh	      &  KUG 1014+433  \\
053 & Id10192117 & 154.7562 & 21.2836 & 0.0036 & 14.56 & 14.34 & 16.21 & 15.94    & 	TS	      &  PGC030133   \\
054 & Id10251708 & 156.2691 & 17.1494 & 0.0025 & 11.54 & 11.30 & 12.29 & 12.87    & 	TS	      &  NGC3239   \\
055 & Id10291610 & 157.4553 & 16.1809 & 0.0108 & 15.12 & 14.89 & 16.96 & 17.18    & 	SL	      &  MRK 0631  \\
056 & Id1034-0221 & 158.5039 & -2.3663 & 0.0067 & 15.15 & 15.13 & 15.63 & 15.95    & 	T	      &  PGC031246   \\
057 & Id10345046 & 158.6911 & 50.7683 & 0.0020 & 14.14 & 13.78 & 15.53 & \nodata       & 	I	      &  UGC 05740  \\
058 & Id10354614 & 158.8002 & 46.2367 & 0.0016 & 16.68 & 16.41 & 18.44 & 17.51    & 	Sh	      &     \\
059 & Id10531646 & 163.3549 & 16.7711 & 0.0035 & 12.58 & 12.33 & 13.69 & 14.18    & 	I	      &  PGC032694   \\
060 & Id10535707 & 163.4561 & 57.1186 & 0.0064 & 13.75 & 13.45 & 15.46 & 14.77    & 	T	      &  NGC 3440  \\
061 & Id10545418 & 163.6635 & 54.3052 & 0.0045 & 11.99 & 11.63 & 13.94 & 13.30    &       I             & NGC3448 \\
062 & Id11011636 & 165.4623 & 16.6069 & 0.0098 & 14.27 & 13.94 & 16.26 & 15.72    & 	F	      &  UGC 06104  \\
063 & Id1109-0258 & 167.4654 & -2.9778 & 0.0172 & 15.19 & 14.89 & 17.05 & 17.64    & 	I	      &  CGCG 011-014  \\
064 & Id11132131 & 168.4562 & 21.5205 & 0.0048 & 14.66 & 14.26 & \nodata & \nodata          & 	I	      &  UGC 06258  \\
065 & Id11200231 & 170.0612 & 2.52246 & 0.0054 & 13.48 & 13.21 & 14.98 & 14.42    & 	I	      &     \\
066 & Id11221319 & 170.6666 & 13.3305 & 0.0137 & 15.13 & 14.88 & 16.27 & 15.59    & 	I	      &  IC 2776  \\
067 & Id11253803 & 171.3825 & 38.0605 & 0.0070 & 14.24 & 14.01 & 15.79 & 15.70    & 	IMSh      &  UGC 06433 \\
068 & Id1125-0039 & 171.4670 & -0.6615 & 0.0187 & 16.44 & 16.17 & 17.76 & 17.20    & 	I	      &  SHOC 324  \\
069 & Id11292034 & 172.3137 & 20.5831 & 0.0047 & 14.02 & 13.72 & 15.70 & 15.53    & 	IMSh      &  	IC 0700  \\
070 & Id11350233 & 173.7706 & 02.5513 & 0.0174 & 15.05 & 14.81 & 16.70 & 16.46    & 	T	      &  UGC 06558  \\
071 & Id11351601 & 173.9550 & 16.0266 & 0.0172 & 16.89 & 16.62 & \nodata & \nodata          & 	IB	      &    \\
072 & Id11401924 & 175.1175 & 19.4097 & 0.0113 & 15.26 & 14.84 & \nodata & \nodata          & 	Sh	      &  KUG 1137+196  \\
073 & Id11414623 & 175.3414 & 46.3932 & 0.0024 & 15.09 & 14.74 & 17.30 & 16.63    & 	E	      &  PGC036272   \\
074 & Id11412457 & 175.3545 & 24.9516 & 0.0113 & 15.26 & 14.83 & 16.84 & 17.66    & 	TE	      &  KUG 1138+252  \\
075 & Id11451711 & 176.4793 & 17.1923 & 0.0110 & 15.82 & 15.71 & 17.54 & 17.41    & 	I	      &  UGC 06741  \\
076 & Id1148-0138 & 177.0757 & -1.6399 & 0.0130 & 15.88 & 15.66 & 16.94 & 16.92    & 	Sh	      &  UM 454  \\
077 & Id11501501 & 177.5113 & 15.0231 & 0.0024 & 14.81 & 14.72 & \nodata & \nodata          & 	Sh	      &  MRK 0750  \\
078 & Id11502557 & 177.5840 & 25.9618 & 0.0125 & 13.90 & 13.57 & 15.89 & 15.40    & 	I	      &  UGC 06806  \\
079 & Id1152-0228 & 178.1549 & -2.4694 & 0.0034 & 14.18 & 14.10 & 15.13 & \nodata       & 	E	      &  UGC 06850  \\
080 & Id11563207 & 179.1355 & 32.1303 & 0.0102 & 15.89 & 15.82 & 16.79 & 16.51    & 	I	      &     \\
081 & Id12002453 & 180.0115 & 24.8892 & 0.0112 & 16.86 & 16.53 & 18.22 & 17.29    & 	E	      &     \\
082 & Id12032526 & 180.9725 & 25.4352 & 0.0107 & 14.00 & 13.72 & 14.82 & 15.25    & 	IM	      &  UGC 07040  \\
083 & Id12065858 & 181.5600 & 58.9711 & 0.0108 & 15.36 & 15.26 & 16.47 & 16.11    & 	I	      &  PGC038384   \\
084 & Id12111929 & 182.9358 & 19.4906 & 0.0116 & 15.03 & 14.62 & \nodata & \nodata          & 	T	      &  PGC038842   \\
085 & Id12131705 & 183.2770 & 17.0988 & 0.0143 & 15.53 & 15.20 & 17.18 & 16.63    & 	TS	      &  MRK 0762  \\
086 & Id12242109 & 186.0920 & 21.1569 & 0.0031 & 14.81 & 14.36 & 16.69 & 17.49    & 	Sh	      &  UGC 07485  \\
087 & Id12241323 & 186.1191 & 13.3858 & 0.0199 & 16.47 & 16.28 & 17.64 & 16.87    &   I        &   VIII Zw 186\\
088 & Id12250548 & 186.4687 & 05.8095 & 0.0050 & 14.83 & 14.54 & 16.63 & 16.18    & 	Sh	      &  VCC0848   \\
089 & Id12284405 & 187.0463 & 44.0935 & 0.0006 & 9.24  & 9.55  & 10.86 & 10.81    &        I             &NGC4449 \\
090 & Id12304138 & 187.6515 & 41.6436 & 0.0018 & 9.68  & 9.35  & 12.00 & 11.44    &        I            &  NGC4490 \\
091 & Id12324937 & 188.0033 & 49.6303 & 0.0145 & 15.11 & 14.89 & 16.65 & 16.25    & 	IM	      &  PGC041500   \\
092 & Id12383805 & 189.7371 & 38.0902 & 0.0074 & 15.24 & 15.00 & 16.42 & 16.07    & 	I	      &  UGC 07816  \\
093 & Id1239-0348 & 189.8345 & -3.8083 & 0.0084 & 15.12 & 14.90 & 16.46 & 15.93    & 	IB	      &  PGC042338   \\
094 & Id12394526 & 189.9053 & 45.4392 & 0.0125 & 15.85 & 15.48 & 16.94 & 16.52    & 	I	      &     \\
095 & Id1241-0007 & 190.4004 & -0.1216 & 0.0158 & 15.45 & 15.07 & 16.40 & 17.25    & 	T	      &  UM 512  \\
096 & Id12444500 & 191.0289 & 45.0050 & 0.0123 & 14.37 & 14.19 & 15.78 & 15.45    & 	IB	      &  PGC042874   \\
097 & Id12464814 & 191.5972 & 48.2352 & 0.0030 & 15.17 & 14.94 & 16.43 & 16.20    & 	Sh	      &  UGCA 297  \\
098 & Id12474709 & 191.8241 & 47.1616 & 0.0196 & 15.58 & 15.07 & 18.04 & 17.19    & 	TE	      &  MRK 0225  \\
099 & Id1249-0434 & 192.4243 & -4.5797 & 0.0047 & 13.84 & 13.64 & 14.77 & 15.10    & 	E	      &  NGC 4678  \\
100 & Id12530427 & 193.3083 & 04.4650 & 0.0024 & 12.92 & 12.62 & 14.19 & 14.61    & 	Sh	      &  NGC 4765  \\
101 & Id12540239 & 193.7166 & 02.6527 & 0.0031 & 13.30 & 13.11 & 14.14 & 14.38    & 	I	      &  NGC 4809/ARP 2\\
102 & Id12561630 & 194.2144 & 16.5067 & 0.0041 & 15.72 & 15.09 & \nodata & \nodata          & 	TE	      &     \\
103 & Id1258-0423 & 194.6983 & -4.3861 & 0.0047 & 16.36 & 16.00 & 16.86 & 17.54    & 	T	      &     \\
104 & Id13161232 & 199.2180 & 12.5482 & 0.0032 & 13.85 & 13.55 & 15.43 & 15.07    & 	Sh	      &  NGC 5058  \\
105 & Id13193015 & 199.9133 & 30.2566 & 0.0071 & 13.41 & 13.03 & 15.92 & 15.38    & 	T	      &  NGC 5089  \\
106 & Id1328-0202 & 202.1943 & -2.0380 & 0.0123 & 14.29 & 13.91 & 15.45 & 15.87    & 	TS	      &  PGC047278   \\
107 & Id13303119 & 202.5723 & 31.3327 & 0.0161 & 14.46 & 14.29 & 15.82 & 15.47    & 	IM	      &  UGC 08496  \\
108 & Id13335449 & 203.2852 & 54.8275 & 0.0176 & 15.49 & 15.19 & \nodata & \nodata          & 	E	      &  PGC047713   \\
109 & Id13343125 & 203.5622 & 31.4250 & 0.0166 & 14.97 & 14.77 & 16.27 & 15.89    & 	I	      &  UGC 08548  \\
110 & Id13425241 & 205.7475 & 52.6883 & 0.0059 & 16.18 & 15.87 & \nodata & \nodata          & 	I	      &  MRK 1481  \\
111 & Id13433644 & 205.8047 & 36.7493 & 0.0197 & 15.64 & 15.28 & 16.75 & 15.84    & 	TS	      &     \\
112 & Id13434311 & 205.8624 & 43.1885 & 0.0083 & 16.18 & 15.98 & 17.82 & 17.17    & 	Sh	      &     \\
113 & Id13493743 & 207.4594 & 37.7306 & 0.0081 & 16.01 & 15.70 & 17.09 & 17.36    & 	IB	      &     \\
114 & Id13516422 & 207.9732 & 64.3728 & 0.0058 & 15.11 & 14.93 & 16.27 & 15.88    & 	IM	      &  PGC049221   \\
115 & Id1355-0600 & 208.9394 & -6.0028 & 0.0066 & 14.48 & 14.21 & 15.25 & 15.68    & 	I	      &  PGC049521   \\
116 & Id1356-0441 & 209.1966 & -4.6923 & 0.0098 & 16.99 & 16.81 & 17.00 & 17.78    & 	T	      &     \\
117 & Id13563656 & 209.2236 & 36.9454 & 0.0197 & 17.59 & 17.45 & 17.07 & 16.06    & 	I	      &     \\
118 & Id14005514 & 210.1351 & 55.2460 & 0.0127 & 16.04 & 15.91 & 16.83 & 16.59    & 	I	      &     \\
119 & Id14010759 & 210.4178 & 07.9979 & 0.0179 & 16.93 & 16.59 & 18.53 & 17.48    & 	I	      &     \\
120 & Id14041243 & 211.2216 & 12.7288 & 0.0136 & 13.81 & 13.60 & 14.20 & 14.35    & 	I	      &  UGC 09002  \\
121 & Id1410-0234 & 212.5531 & -2.5744 & 0.0051 & 13.44 & 13.15 & 15.34 & 15.18    & 	IM	      &  UGC 09057  \\
122 & Id14182530 & 214.6066 & 25.5018 & 0.0149 & 15.25 & 14.92 & \nodata & \nodata          & 	LS	      &  PGC051103   \\
123 & Id14182149 & 214.6805 & 21.8175 & 0.0085 & 14.66 & 14.71 & 15.79 & 15.53    & 	IM	      &  PGC051120   \\
124 & Id1421-0345 & 215.3427 & -3.7588 & 0.0091 & 14.44 & 14.24 & \nodata & 15.59       & 	LS	      &  PGC051291   \\
125 & Id14294426 & 217.4622 & 44.4476 & 0.0092 & 14.46 & 14.24 & 16.37 & 16.00    & 	IM	      &  PGC051798   \\
126 & Id14312714 & 217.7876 & 27.2373 & 0.0150 & 14.41 & 14.26 & \nodata & \nodata          & 	I	      &  MRK 0685  \\
127 & Id14362827 & 219.0358 & 28.4505 & 0.0063 & 15.43 & 15.14 & 16.75 & 16.43    & 	Sh	      &  HARO 43  \\
128 & Id14365127 & 219.1908 & 51.4597 & 0.0078 & 15.56 & 15.13 & 17.38 & 16.87    & 	E	      &  PGC052226   \\
129 & Id14392323 & 219.9387 & 23.3965 & 0.0150 & 15.73 & 15.60 & 17.38 & 16.74    & 	I	      &  UGC 09450  \\
130 & Id14453124 & 221.3852 & 31.4155 & 0.0049 & 14.65 & 14.56 & 15.38 & 14.97    & 	I	      &  UGC 09506  \\
131 & Id1448-0342 & 222.2000 & -3.7163 & 0.0031 & 14.24 & 13.98 & 14.96 & 15.26    & 	I	      &  PGC052893   \\
132 & Id14493623 & 222.4531 & 36.3965 & 0.0062 & 16.33 & 16.22 & 17.16 & 16.90    & 	I	      &     \\
133 & Id14503534 & 222.7356 & 35.5721 & 0.0039 & 14.41 & 14.29 & 15.12 & 15.00    & 	I	      &  UGC 09560  \\
134 & Id14543012 & 223.5488 & 30.2095 & 0.0094 & 14.72 & 14.60 & 15.87 & 15.65    & 	A	      &  UGC 09588  \\
135 & Id14572640 & 224.4108 & 26.6683 & 0.0042 & 15.41 & 15.15 & 16.88 & 17.21    & 	S	      &     \\
136 & Id15052341 & 226.3632 & 23.6883 & 0.0162 & 15.35 & 15.09 & 17.45 & 16.81    & 	A	      &  UGC 09698  \\
137 & Id1507-0239 & 226.7837 & -2.6627 & 0.0068 & 16.21 & 15.96 & 17.33 & 17.27    & 	I	      &     \\
138 & Id15075511 & 226.9514 & 55.1857 & 0.0111 & 13.99 & 13.63 & 21.70 & 19.67    & 	Sh	      &  UGC 09737  \\
139 & Id15091950 & 227.3164 & 19.8486 & 0.0158 & 16.99 & 16.58 & 18.42 & 17.69    & 	I	      &     \\
140 & Id15174257 & 229.3553 & 42.9559 & 0.0178 & 15.60 & 15.23 & 17.15 & 16.65    & 	IM	      &  PGC054571   \\
141 & Id15182205 & 229.6666 & 22.0863 & 0.0158 & 15.50 & 15.11 & 17.09 & 17.47    & 	T	      &  PGC054647   \\
142 & Id15271117 & 231.8578 & 11.2842 & 0.0129 & 16.14 & 15.75 & \nodata & \nodata          & 	TS	      &     \\
143 & Id15271258 & 231.9358 & 12.9708 & 0.0128 & 18.63 & 18.40 & 17.74 & 18.54    & 	I	      &     \\
144 & Id15292600 & 232.3709 & 26.0075 & 0.0067 & 14.97 & 14.86 & \nodata & \nodata          & 	I	      &     \\
145 & Id15354648 & 233.7504 & 46.8146 & 0.0188 & 15.58 & 15.39 & \nodata & 16.54       & 	T	      &  I Zw 116 \\
146 & Id15353840 & 233.9737 & 38.6777 & 0.0186 & 14.14 & 13.95 & 15.96 & 15.53    &   I        &  \\
147 & Id15363040 & 234.0806 & 30.6811 & 0.0058 & 14.84 & 14.59 & \nodata & \nodata          & 	I	      &  PGC055576   \\
148 & Id15480414 & 237.0172 & 04.2423 & 0.0131 & 16.73 & 16.47 & \nodata & \nodata          & 	T	      &     \\
149 & Id15511001 & 237.7562 & 10.0311 & 0.0146 & 15.88 & 15.62 & 16.28 & 17.02    & 	T	      &  CGCG 078-083  \\
150 & Id15541637 & 238.6718 & 16.6173 & 0.0079 & 14.57 & 14.14 & \nodata & \nodata          & 	I	      &  UGC 10086  \\
151 & Id16055045 & 241.4178 & 50.7544 & 0.0128 & 15.66 & 15.51 & 16.97 & 16.49    & 	I	      &     \\
152 & Id16054119 & 241.4459 & 41.3182 & 0.0066 & 13.14 & 12.94 & 14.60 & 14.29    &    I         & UGC 10200 \\
153 & Id16060634 & 241.6708 & 06.5808 & 0.0058 & 14.69 & 14.33 & 16.10 & 15.57    & 	Sh	      &  PGC057169   \\
154 & Id16212838 & 245.3675 & 28.6399 & 0.0029 & 14.40 & 14.05 & 16.01 & 16.31    & 	Sh	      &  UGC 10351  \\
155 & Id16274825 & 246.9728 & 48.4248 & 0.0134 & 16.47 & 16.19 & 17.95 & 17.29    & 	Sh	      &     \\
156 & Id16312024 & 247.9531 & 20.4107 & 0.0171 & 14.54 & 14.36 & 15.97 & 15.44    & 	Sh	      &  MRK 0884  \\
157 & Id14503534 & 249.5121 & 26.4527 & 0.0144 & 15.41 & 15.36 & 16.44 & 16.07    & 	E	      &     \\
158 & Id16472105 & 251.7956 & 21.0952 & 0.0090 & 15.42 & 15.35 & \nodata & \nodata          & 	IM	      &     \\
159 & Id17135919 & 258.2862 & 59.3277 & 0.0036 & 13.98 & 13.76 & 15.00 & 14.72    &   IM       &   UGC10770\\
160 & Id2119-0733 & 319.9287 & -7.5523 & 0.0090 & 14.11 & 13.86 & 15.87 & 15.59    & 	I	      &  PGC066559   \\
161 & Id21421518 & 325.6345 & 15.3000 & 0.0122 & 15.38 & 15.15 & 15.85 & 16.28    & 	T	      &  AGC 748645  \\
162 & Id22021945 & 330.6332 & 19.7501 & 0.0054 & 13.54 & 13.21 & 15.28 & 14.81    &   IM       &   IC 1420\\
163 & Id22080441 & 332.0383 & 4.69000 & 0.0135 & 14.18 & 13.98 & 15.94 & 15.26    &   I        &   PGC068112\\
164 & Id22162255 & 334.0320 & 22.9333 & 0.0128 & 14.76 & 14.46 & 15.79 & 16.32    & 	Sh	      &  KUG 2213+226  \\
165 & Id22271205 & 336.8610 & 12.0944 & 0.0118 & 15.98 & 15.65 & 17.30 & 17.65    & 	Sh	      &     \\
166 & Id22391352 & 339.8411 & 13.8822 & 0.0173 & 15.08 & 14.81 & 16.77 & 16.22    & 	Sh	      &     \\
167 & Id23021636 & 345.7469 & 16.6052 & 0.0069 & 13.47 & 13.18 & \nodata & \nodata          &   T        &   NGC 7468\\
168 & Id2319-0059 & 349.9917 & -0.9855 & 0.0121 & 16.12 & 15.81 & 18.01 & 17.91    & 	TS	      &     \\
169 & Id2320-0052 & 350.1466 & -0.8809 & 0.0145 & 16.41 & 16.20 & \nodata & \nodata          & 	I	      &  UM 158  \\
170 & Id2324-0006 & 351.0990 & -0.1075 & 0.0090 & 14.22 & 14.05 & 15.49 & 15.30    & 	I	      &  UGC 12578  \\
171 & Id23260157 & 351.6245 & 01.9602 & 0.0172 & 15.37 & 15.08 & 16.58 & 17.27    & 	Sh	      &  CGCG 380-056  \\
172 & Id23261144 & 351.6595 & 11.7423 & 0.0125 & 14.76 & 14.45 & \nodata & \nodata          &   IM       &   KUG 2324+114\\
173 & Id23302531 & 352.5412 & 25.5327 & 0.0191 & 14.54 & 14.27 & \nodata & 15.38       &   IM       &   III Zw 107\\
174 & Id23312856 & 352.9928 & 28.9472 & 0.0182 & 14.74 & 14.77 & 15.79 & 15.33    &   IM       &   MRK  0930\\
175 & Id23371759 & 354.3894 & 17.9962 & 0.0084 & 13.79 & 13.43 & \nodata & \nodata          &   IM       &   UGC12710\\
176 & Id2340-0053 & 355.1846 & -0.8874 & 0.0191 & 17.41 & 17.17 & 18.06 & 17.19    & 	IM	      &    \\
177 & Id23591448 & 359.9042 & 14.8078 & 0.0058 & 13.21 & 12.82 & 14.53 & 14.05    & 	Sh	      &  NGC 7800  \\             
\enddata
\tablecomments{The first column is number. We list the Interacting dwarf (Id), coordinates (RA and Dec) and redshift in columns 2, 3, 4, and 5, respectively. The Ids are in `hmdm' format. FUV, NUV, g, and r photometric data is listed in columns 6-9. We present morphological class of merging dwarf systems, obtained according to $\S$ \ref{class} in column 10.  In the last column we provide Name of galaxies that we found in NED. } 
\end{deluxetable*}

\begin{itemize}
\item Interacting (I): In this class, we identify ongoing interactions between two dwarf galaxies. If the two interacting dwarf galaxies are visibly distinct, we simply designate it with an `I' (e.g. Id0217-0742), and if they are overlapping, or the progenitor galaxies are not distinct, we also give it an `M' (Merged, e.g Id01250759). Additionally, if we see a bridge connecting the interacting galaxies we add `B' (for bridge, e.g. Id01482838). A dwarf analog of the famous Antennae system (NGC 4038/NGC 4039) is represented by `A' (for Antennae e.g. Id0202-0922). 
\item Shell (Sh): The presence of shell features can be seen e.g. Id0155-0011.
\item Tidal tail (T): Simply defined as the presence of amorphous tidal features, mostly tidal streams or plumes, which can not be placed into the above classifications, e.g Id08092137. We notice that the majority of tidal tails are relatively redder than their galaxy's main body (so likely a distinct stellar population). Thus, they might better be described as stellar streams, in which case we add an `S', e.g. Id0222-0830.  Also, if we see a loop of a stellar stream around the galaxies, we identify this with an `L', e.g. Id09530702.
\end{itemize}

 We show various examples of these classification in Figure \ref{mclass}. It is worth noting that the above classification scheme is not mutually exclusive, and in a number of cases there are overlaps. For example, some interacting galaxies also posses multiple tidal features, like shells or stellar stream, even when the two parent dwarf galaxies are not yet fully merged. Id11253803 is the best example of this kind. We show an example of these different morphological classes of merging dwarf galaxies in Figure \ref{mclass}. 

\section{Data analysis}\label{data}
To perform the photometric analysis and measure the total luminosity, we exclusively use the SDSS image data, unless explicitly mentioned otherwise.
This is because the SDSS provides the best homogeneous imaging data. We retrieved archival images from the SDSS-III database \citep{Abazajian09}. Since the SDSS data archive provides well calibrated and sky-background subtracted images, no further effort has been made in this regard. We derive the $g$ and $r-$band magnitudes. To do this, we measure the total flux by placing a large aperture which covers both interacting galaxies and the stellar streams around them. While doing so, unrelated background and foreground objects were masked manually. This procedure is quite straight forward if the interacting galaxies are not well separated or already merged. In the case of interacting systems, when the galaxies involved are well separated (class I), the apertures are chosen in two different ways --first a large aperture covering both the interacting galaxies is used to measure the total flux of the system, as done for the other classes.  Additionally, we also use smaller apertures to measure the flux of the individual galaxies. However, we emphasize that we only use the aperture photometry of the individual interacting galaxies to calculate their mass ratio. For the rest of the physical parameters we present in this work, values are given for the total system (e.g. magnitudes, $g-r$ colors, stellar masses and star-formation rates).

There are only six candidate galaxies which are located outside of the SDSS covered region of sky. In these cases, we use images from the Legacy survey. We maintain similar procedures for the aperture photometry as were applied with the SDSS images.

\begin{deluxetable*}{cccRccCccc}
\tablenum{2}
\tablecaption{Derived properties of merging dwarf galaxies \label{ptb}}
\tablecolumns{10}
\tablehead{
\colhead{Number} &
\colhead{Distance} &
\colhead{g-r} &
\colhead{M$_{B}$} &
\colhead{M$_{*}$ } &
\colhead{M1:M2} &
\colhead{SFR} &
\colhead{M$_{HI}$} &
\colhead{Set} &
\colhead{no. neighbor} \\
\colhead{} &
\colhead{Mpc } &
\colhead{mag } &
\colhead{mag} &
\colhead{log(M$_{\sun}$)} &
\colhead{} &
\colhead{log(M$_{\sun}$/yr)} &
\colhead{log(M$_{\sun}$)}  &
\colhead{} &
\colhead{}
}
\colnumbers
\startdata
001 &  16.52 & 0.11 &  -14.96   & 7.81  &  5      &  -1.51   & 8.49   & 1    & 22\\
002 &  41.26 & 0.04 &  -17.66   & 8.78  &  2      &  -0.35   & 9.30   & 0    & 9 \\
003 &  53.29 & 0.43 &  -17.49   & 9.35  &  4      &  -0.68   & 8.76   & 1    & 3 \\
004 &  51.57 & 0.46 &  -16.72   & 9.09  &  \nodata      &  -0.91   & 8.43   & 0    & 3 \\
005 &  77.06 & 0.30 &  -18.68   & 9.61  &  \nodata      &  -0.40   & \nodata   & 0    & 1 \\
........  &  ........ & ........ &  ..........   & .......  &  ...      &  ........   & .....  & ..    & .. \\
\enddata
\tablecomments{A portion of the Table is shown here for guidance regarding its contents and form. The table in its entirety will be published as part of the online catalog. \\
Col.(1): Number, Col.(2): Adopted distance to the galaxy, Col.(3): g-r color, Col(4): B-band absolute magnitude, Col(5): Stellar mass, Col(6): Mass ratio of interacting galaxies, Col(7): Star formation rate, Col(8) HI mass: , Col(9): Satellite or not -- 1 for yes and 0 for no, Col(10): Number of neighboring galaxies within our search criteria -- see text \S\ref{data}.}
\end{deluxetable*}

For many galaxies (146 out of 177), we found there were GALEX all-sky survey observations available \citep{Martin05}. Since they are mostly star-forming, almost all are detected in FUV and NUV-band GALEX all-sky survey images. In these cases, we perform aperture photometry on the GALEX image, following the same procedure as we used for the optical images. However, we only calculate the total UV flux of the systems, and not for the individual galaxies, because the GALEX images have a spatial resolution of only 5" and the individual galaxies are not well resolved.

The distances to the galaxies are taken from NED. For those where NED does not provide a redshift independent distance, we calculate it based on Hubble flow assuming the cosmological parameters defined in \S\ref{intro}. We use the python code, {\sc cosmocalc}, available in {\sc astropy} to calculate cosmological distances based on the radial velocities. The radial velocities are not corrected for Virgo-centric flow.

\begin{figure}
\includegraphics[width = 8cm]{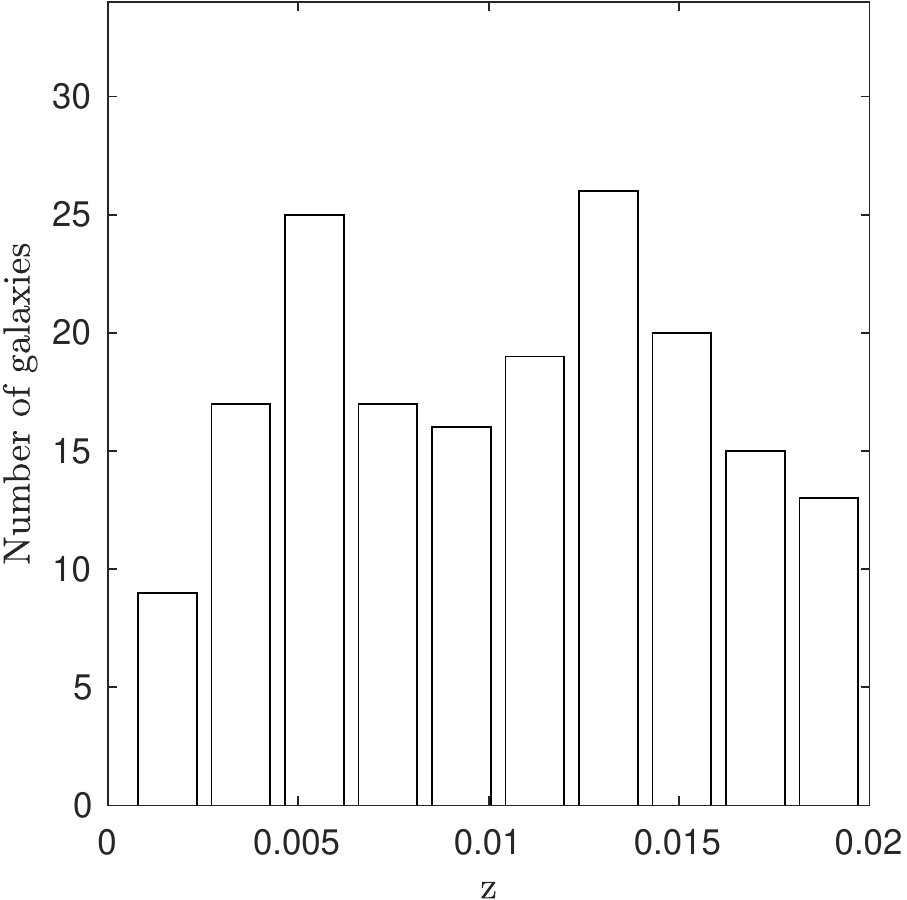}
\caption{Redshift distribution of the sample}
\label{zhist}
\end{figure}
 
 \begin{figure}
\includegraphics[width = 8cm]{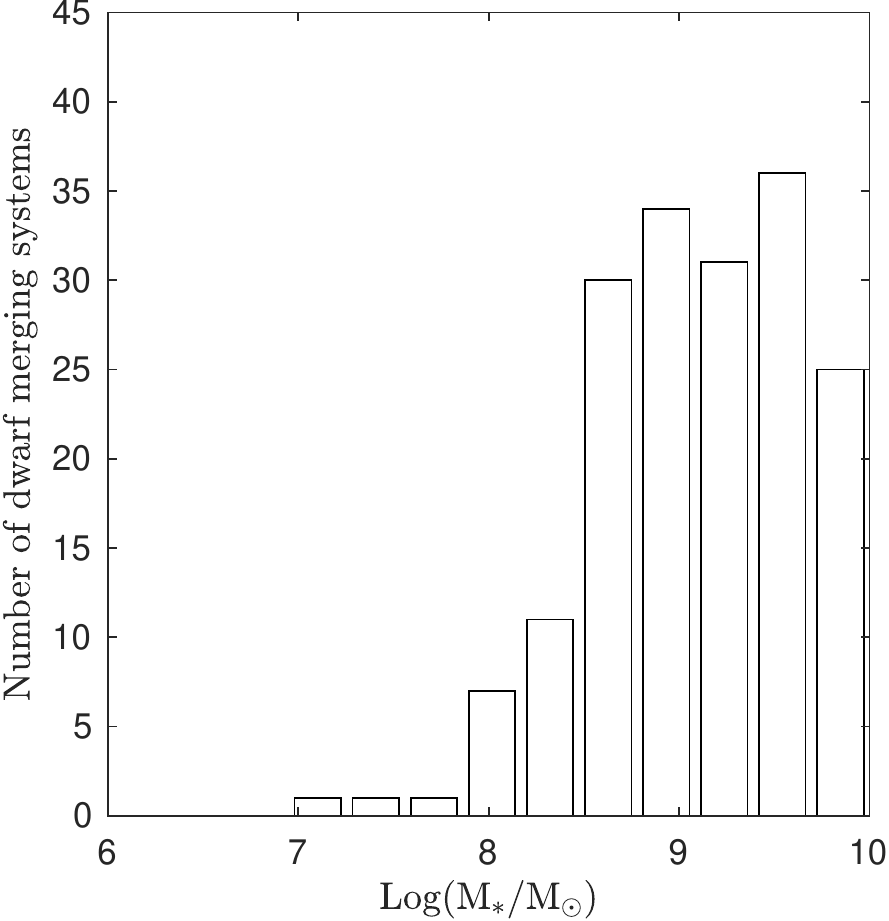}
\caption{Distribution of the logarithm of the stellar mass of merging dwarf systems.}
\label{stm}
\end{figure}

\begin{figure}
\includegraphics[width = 8.5cm]{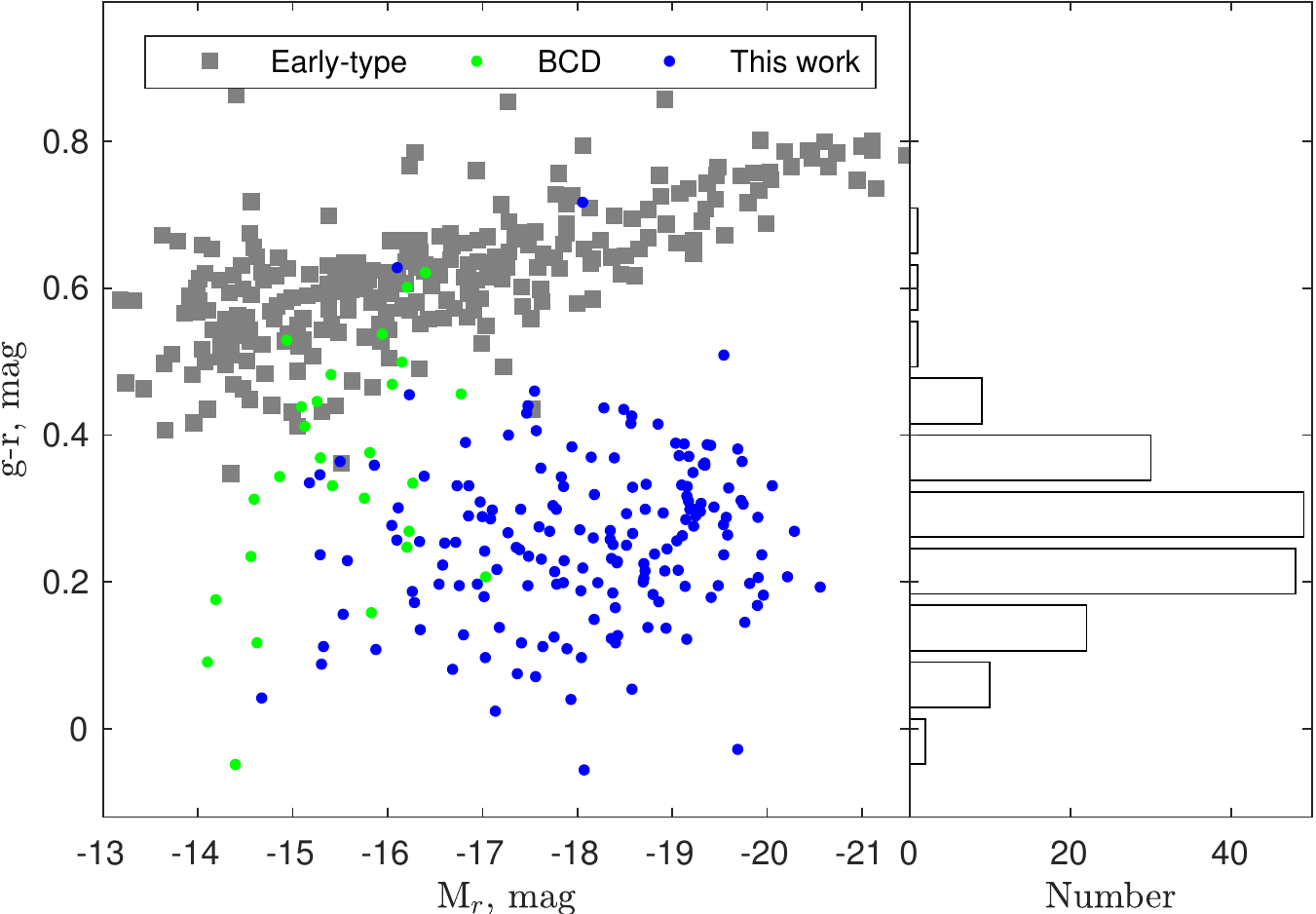}
\caption{The optical color-magnitude relation. Blue dots represent interacting dwarfs. The comparison sample are early-type galaxies (gray square) and BCDs (green dots) taken from \cite{Janz09} and \cite{Meyer14}, respectively.}
\label{colhist}
\end{figure}

\begin{figure}
\includegraphics[width = 8cm]{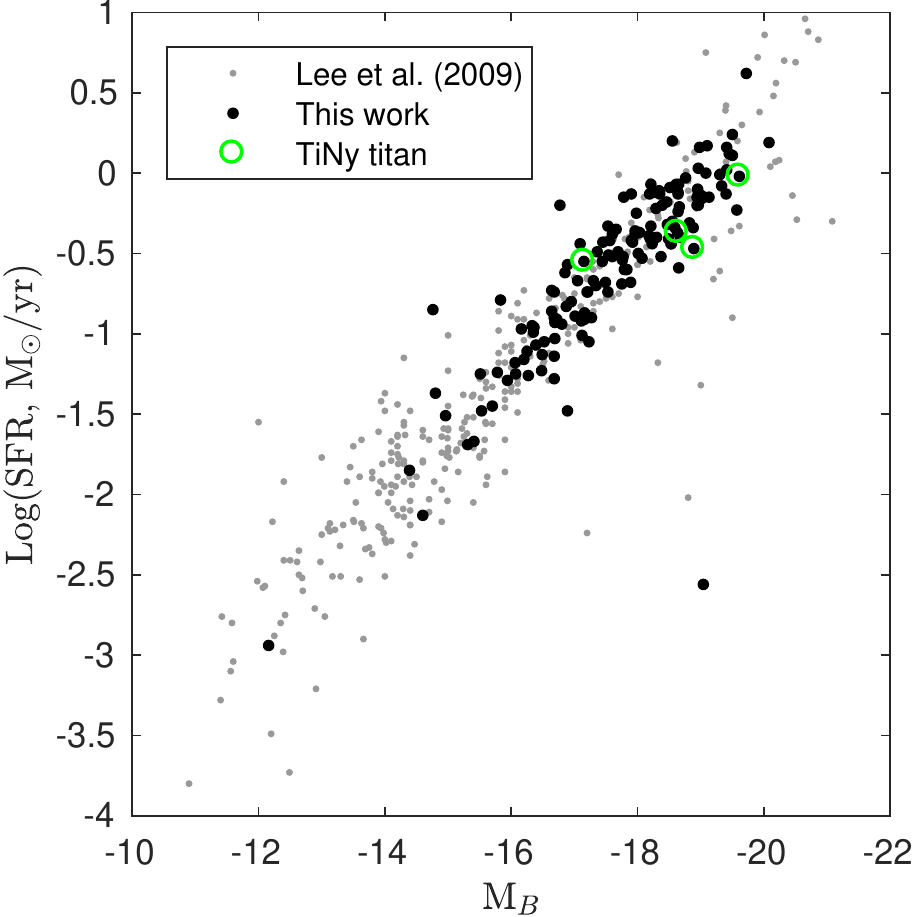}
\caption{Relation between star-formation rate versus blue-band absolute magnitude. The black symbol represents merging dwarf systems and gray symbols are the Lee et al (2009) galaxies. Those interacting pairs that are found in both our sample and those of the Tiny titan sample are shown with green circles. }
\label{sfr}
\end{figure}

\begin{figure}
\includegraphics[width = 8.5cm]{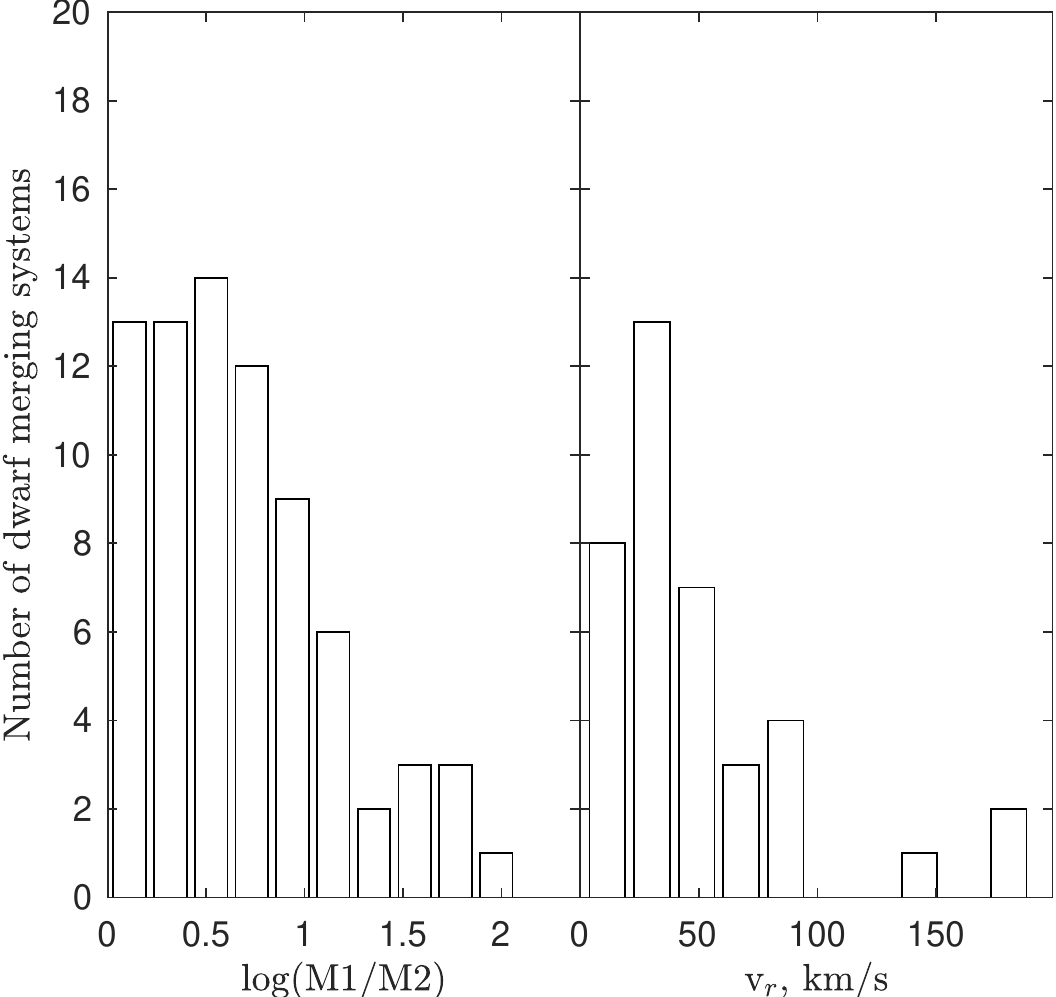}
\caption{Distribution of mass ratio and relative line of sight velocity of interacting dwarf pairs. Each panel contains different numbers of galaxies; for the mass ratio there are 76 and for the velocity separation there are 38, for the reasons given in the text, see \S\ref{result}}
\label{rhist}
\end{figure}

The derived magnitudes were corrected for the Galactic extinction using \cite{Schlafly11}, but not for internal extinction. The star formation rates (SFRs) are derived from the FUV fluxes applying a foreground Galactic extinction correction \cite[A$_{FUV}$ = 7.9 $\times$ E(B -V)][]{Lee09}.  We use the equation \cite[SFR($M_{\sun}$ yr$^{-1}$ ) = 1.4 $\times$10$^{-28}$ L$_{\nu}$(UV)(erg s$^{-1}$ Hz$^{-1}$ )][]{Kennicutt98}.  The stellar masses were derived from the SDSS$-r$ band magnitude with a mass to light ratio tabulated by \cite{Bell03} appropriate to the observed  $g-r$ color.

\section{results}\label{result}
Our morphological classification reveals that there are 98 interacting dwarf galaxy systems. Among these, 22 are classified `Interacting Merger' (IM) type where the boundary between the interacting galaxies can no longer be clearly identified. 30 possess shell features and the rest (49) show tidal tails of different forms. The shell features are mainly found outside of the main body of the galaxies. Some of these resemble the dwarfs with the symmetrical-shaped shell features that were found in \cite{Paudel17} (e.g. Id09381942,  Id10354614, Id12464814). In \cite{Paudel17}, we studied three dwarf galaxies and, with help of idealized numerical simulation, found that they had suffered a very recent (in last few hundred Myr), near equal mass merger which explained their symmetry. However, in some cases, the shell dwarfs do not show such symmetry in their shells (e.g.  Id11253803 ) and in two we find that shell and tidal tails features coexist with each other (Id11253803  and Id11292034). In these cases, the shells are generally higher surface brightness than the tidal tails.

There are three dwarf galaxy systems (Id0202-0922, Id1448-0342, Id14503534) which can be considered dwarf analogues to the Antennae system (NGC 4038/4039).

We present the result of aperture photometry in Table \ref{mtb}. We list the positions (RA and DEC) and redshift of candidate dwarf galaxies in column 2, 3 and 4, respectively. Optical $g$ and $r$ band magnitude are listed in column 5 and 6, respectively. Next we list FUV and NUV band magnitudes in the column 7 and 8, respectively. The classification of morphological feature are given in column 9.

We show the redshift distribution of our catalog of dwarf galaxies in Figure \ref{zhist}. The median redshift of this sample is 0.01. Next, we show the total stellar mass distribution of interacting/merging dwarf galaxies in Figure \ref{stm}.  It is not surprising that this sample is some what biased towards the brighter end of our stellar mass cut. Nevertheless, the range of stellar mass coverage is of order 3 magnitudes, with the median value of log(M$_{*}$/M$_{\sun}$) = 9.1. The minimum mass galaxy, Id10354614, has a similar stellar mass to the local group Fornax dwarf galaxy or Virgo cluster dwarf galaxy VCC1407, both are well known for their shell feature and well discussed as a merger remnant \citep{Paudel17,Coleman04}.

\begin{figure}
\includegraphics[width = 8.5cm]{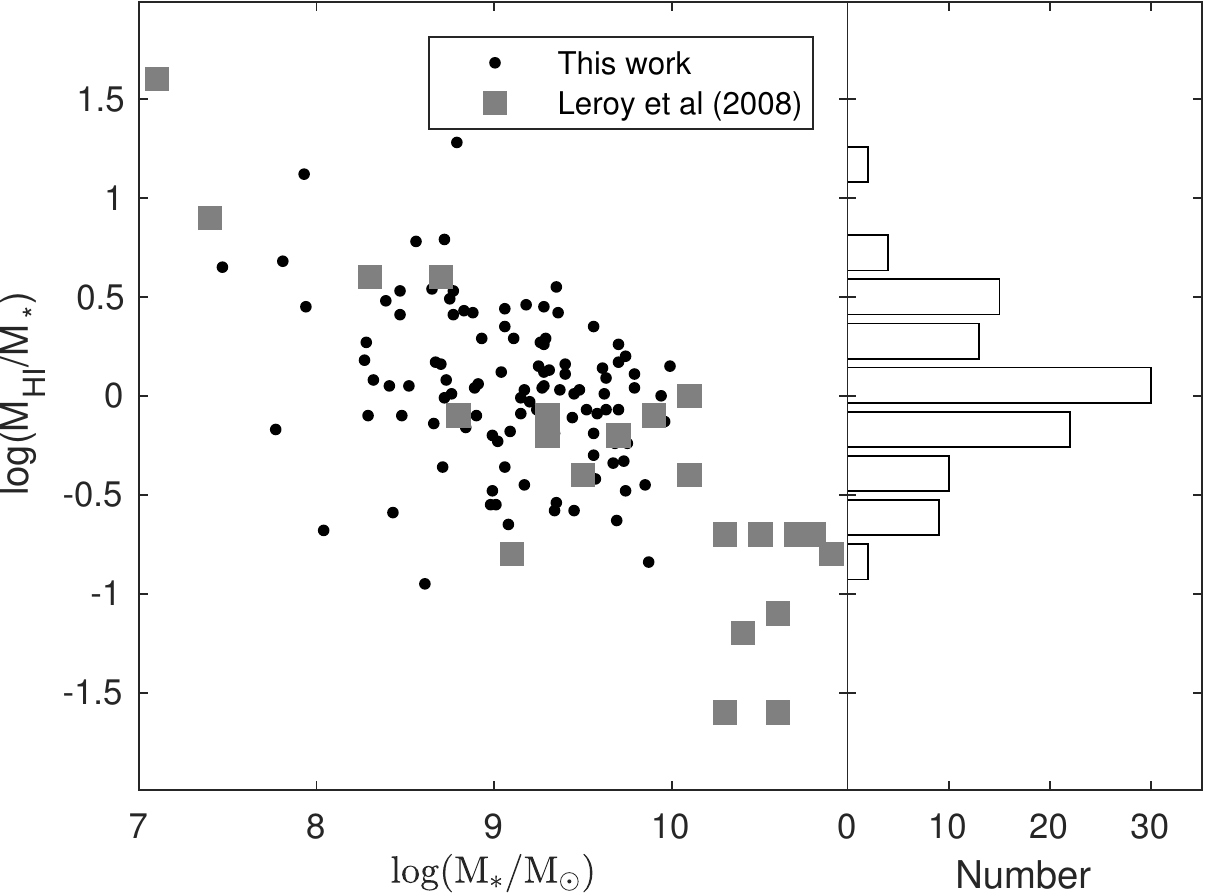}
\caption{Relation between gas mass fraction and stellar mass. The comparison data is from \cite{Leroy08}.}
\label{hihist}
\end{figure}

The $g-r$ color distribution shown in Figure \ref{colhist} right panel, reveals that this sample is overwhelmingly dominated by star-forming galaxies with similar colors to Blue Compact Dwarf galaxies \citep[BCDs][]{Meyer14}. Our sample has a median value of $g-r$ color index = 0.32 mag. For comparison, we also show a sample of early-type galaxies from \cite{Janz09} which clearly offsets from our sample galaxies, creating a red-sequence above the star-forming galaxies in the color-magnitude relation. In fact, there are only three galaxies (Id10080227, Id12474709 and Id12561630) which have a $g-r$ color index more red than 0.5 mag and they are also morphologically akin to the early-type galaxies.

As previously mentioned, the overwhelming majority of galaxies in this sample are blue and they are also detected in the GALEX all sky survey FUV band image which further confirms ongoing active star-formation. Figure \ref{setsfr} illustrates the relation between the B-band absolute magnitude and the star-formation rate. The B-band magnitudes are derived from the SDSS g and r-band magnitudes using the equation $B = g + 0.3130\times(g - r) + 0.2271$ (Lupton 2005\footnote{http://www.sdss3.org/dr8/algorithms/sdssUBVRITransform.php}). For comparison, we also plot data from \cite{Lee09}, see gray symbol, who study the FUV-derived star formation rates of local volume ($<$11 Mpc) star-forming galaxies. From this figure, it is clear that these interacting dwarfs galaxies do not differ from the trend made by local volume star-forming galaxies.

Among the interacting system there are 76 for which we can clearly separate out the individual interacting members (only `I' class), which we will refer to as an `interacting dwarf pair'. To measure the mass ratio of interacting dwarf pair, we perform the aperture photometry in the SDSS r-band images on the individual interacting galaxies. Note that this is actually the flux ratio (larger flux /smaller flux) but, under the assumption of similar stellar populations in both galaxies, we simply use the the term mass ratio. Among the interacting dwarf pairs, we find that both member galaxies share similar $g-r$ colors which also validates our assumption of similar stellar populations. We show the distribution of their mass ratios in Figure \ref{rhist}. It is clear that the majority of interactions are major interaction with a mass ratio of 5 or less, and the median is 4. The first bin of the histogram includes 13 systems (17\% of the total) which can be considered equal mass mergers. In reverse, there are 15 systems (20\% of the total) which have a mass ratio larger than 10, which can be considered a minor merger. The maximum mass ratio is 120 in the case of Id14392323. Among the 76 dwarf interacting pair we find that there are 38 systems where radial velocities are available for both members of interacting pair dwarf galaxies. As the right panel of Figure \ref{rhist} indicates, the relative line of sight velocity between the interacting dwarf pairs is relatively low and, only in two cases, it is higher than 100 km/s.

For our sample of merging dwarf galaxies, we also collected neutral Hydrogen (HI) masses from the CDS server\footnote{http://cdsportal.u-strasbg.fr}. Since this data is assembled from various sources in the literature, we caution about the heterogeneity of the results. The various sources may use different beam sizes, and exposure times, depending on the aim and scope of their individual projects \citep{Paturel03,Meyer04,Giovanelli05,Courtois15}. They are mostly from single dish observations, and we expect that a typical beam size of 3\arcmin, like the Arecibo telescope, would be sufficient to entirely cover the interacting dwarf galaxies, and therefore must be considered as measurements for the total, combined system. We found HI masses for 109 merging dwarf galaxies, as listed in Table \ref{ptb}.

Figure \ref{hihist} reveals the relation between the HI mass fraction and  stellar mass of the star-forming galaxies.  It is clear from this figure that our interacting dwarf sample clearly follow the HI mass fraction and stellar mass relation of other star-forming galaxies in the local Universe \cite{Leroy08}.
We show the distribution of HI mass fraction in the right panel. The median value of the gas mass fraction of our sample is M$_{HI}$/M$_{*}$ = 1.09. 

\section{Discussion}\label{discu}

In this paper, we present a sample of interacting dwarf galaxy systems. Given the large heterogeneity in the data collection procedure, probably part of the scientific discussion can only be considered as a qualitative. However, merging/interacting dwarf galaxies are not thought to be a common phenomenon in the local Universe. According to hierarchical cosmology, theory predicts that they are common in the early-universe. To date, no systematic effort has been made to present a sample of interacting dwarf galaxies which is statistical enough to study the properties of interacting dwarf galaxies and their role in the evolution of low mass galaxies. This is the first publicly available catalog in this regard.

\subsection{Comparison to previous study}\label{compr}
A previous study of interacting pairs of dwarf galaxies, \cite[][here after S15]{Stierwalt15}, mainly focuses on a statistical analysis of environmental effects on interacting pairs of dwarf galaxies \citep{Patton13}. Like in this study, S15 also uses SDSS imaging to select their sample galaxies, therefore we expect that both samples cover the same areas of sky. But probably, the main difference is their redshift coverage. Our sample's redshift range is $<$0.02, while the S15 sample galaxies have redshifts up to 0.07.

In addition to this, S15 performed a careful selection of a control sample and a working sample to remove biases due to the sample selection procedure, when comparing the samples. In contrast, in this work we first aim to present a large catalog of merging dwarf systems which will be helpful for a detailed study of various properties of interacting/merging dwarf galaxies in the future. We provide basic properties, such as sky-position, redshift, stellar-mass and star-formation rate. Further, having these properties in hand we also try to assess the effect of environment on our sample galaxies, comparing gas-mass fraction and star formation rate (SFR) between merging dwarf systems and that of normal galaxies from local volume. We mainly compile our comparison sample data from the literature, thus we caution that our comparative study may not be as statistically rigorous as that of the S15 comparative study between interacting dwarf and non-interacting dwarf galaxies. However, we include the comparison simply to give the properties of our sample some context in comparison to a sample of non-interacting dwarfs of similar mass.

 In S15 sample, the pair galaxies needed to have a separation velocity less than 300 km/s which means they required that there be a measured radial velocity for both galaxies. In contrast, we select interacting dwarf galaxies according to their observed tidal features, and it is not necessary to have a radial velocity for both interacting members. This means we are able to study merging dwarfs over a far greater range of merging stages, even when one dwarf has fully merged with another and the only indication of the event might be the remaining tidal features. A good example of this can be found in our shell feature dwarfs.

While comparing S15 sample with only interacting pair (I class), we find significance difference in mass ratio of member dwarf galaxies of interacting pair. S15 overwhelmingly dominated by small mass ratio pairs, i.e 93$\%$ of their sample is less than mass ratio 5 and in our case less than half, only 42\% , interacting pair have mass ratio less than 5. In addition, while comparing radial velocity separation between interacting, although we find a relatively low number of systems that have radial velocity measurements for both interacting member dwarf galaxies of our sample, we find a clear difference with S15 -- only 2 out of 36 (5\%) have a relative line of sight velocity larger than 100 km/s and 15 out of 60 (25\%) interacting dwarf pairs in the S15 sample have relative line of sight velocities larger than 100 km/s.

Another interesting difference is that S15 find there is an enhanced SFR between dwarf galaxies at small separations from their partner, compared to a control sample of isolated dwarf galaxies. However, in Figure \ref{sfr} we find no evidence for an enhanced SFR in our merging dwarf systems compared to a sample of star-forming galaxies of local volume. One reason we see no clear enhancement in SFR could be because we don't attempt to control for separation distance. Also, S15 compared a homogeneously selected control sample with interacting dwarf-pairs, while we simply use data compiled from the literature as a comparison sample. In fact, a small number of the S15 galaxies can be found in common with this sample, although they follow the same trend as our sample (see Figure \ref{sfr}.)

Another part of the difference could emerge from the way we derived SFR. S15 used catalog values of SFRs from \cite{Brinchmann04}, which is derived from H$_{\alpha}$ emission line flux of the SDSS fiber spectroscopic data. On the other hand, we have used the FUV flux to derive the SFR where the FUV emission traces recent star formation over longer time scales compared to H$_{\alpha}$. However note that, to derive SFR we have used FUV flux only corrected for foreground Galactic extinction but not internal extinction therefore these values are, in many case, would be a lower limit. In the future, we will consider full SED fitting, including infrared wavelengths, in order to better constrain their SFRs.

\subsection{Environment}\label{env}
\begin{figure}
\includegraphics[width = 8.5cm]{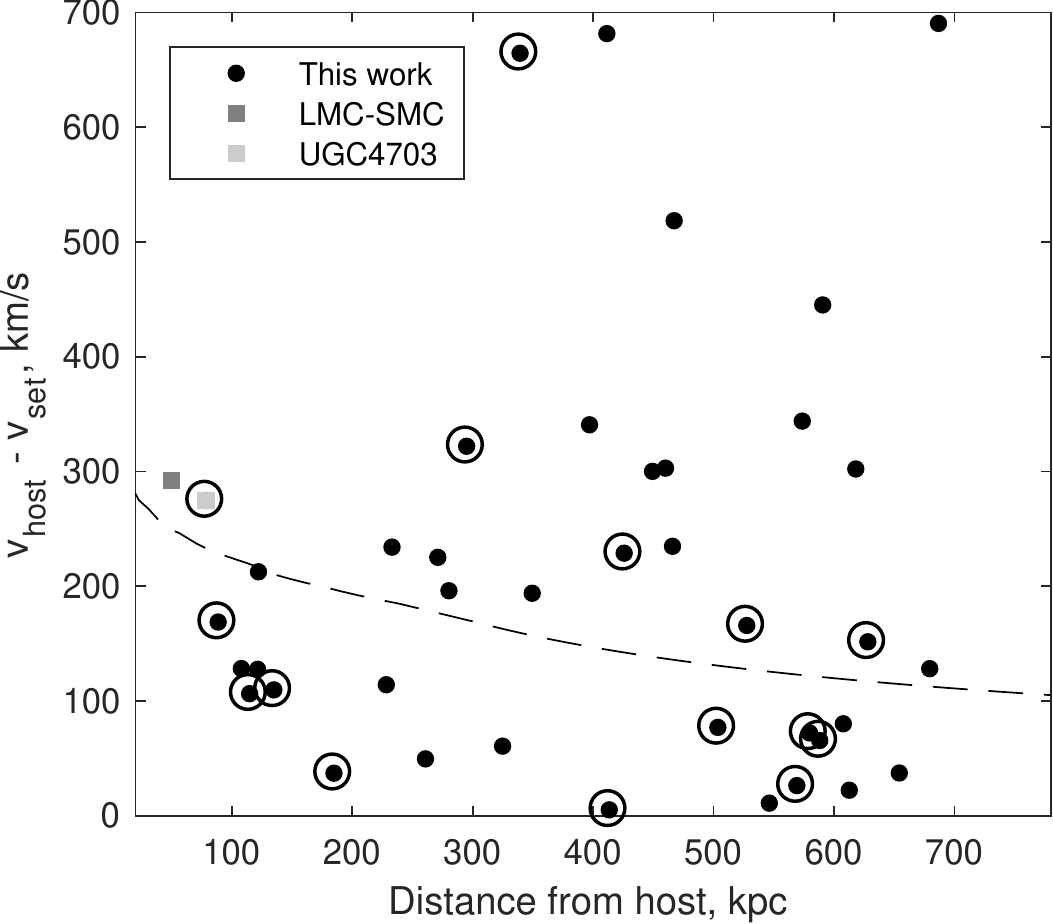}
\caption{Phase-space diagram of merging satellites. Y-axis is relative line of sight velocity between dwarf-merging system and nearby giant galaxy and X-axis is sky-projected physical distance between them. The dashed line represents the escape velocity as a function of radius for a Milky-Way like galaxy, derived from the best match model in \cite{Klypin02}.  We show dwarf interacting pair with a circle. We also show the position of the LMC-SMC pair and UGC 4703 in such a diagram with gray squares. }
\label{setprop}
\end{figure}

\begin{figure}
\includegraphics[width = 8cm]{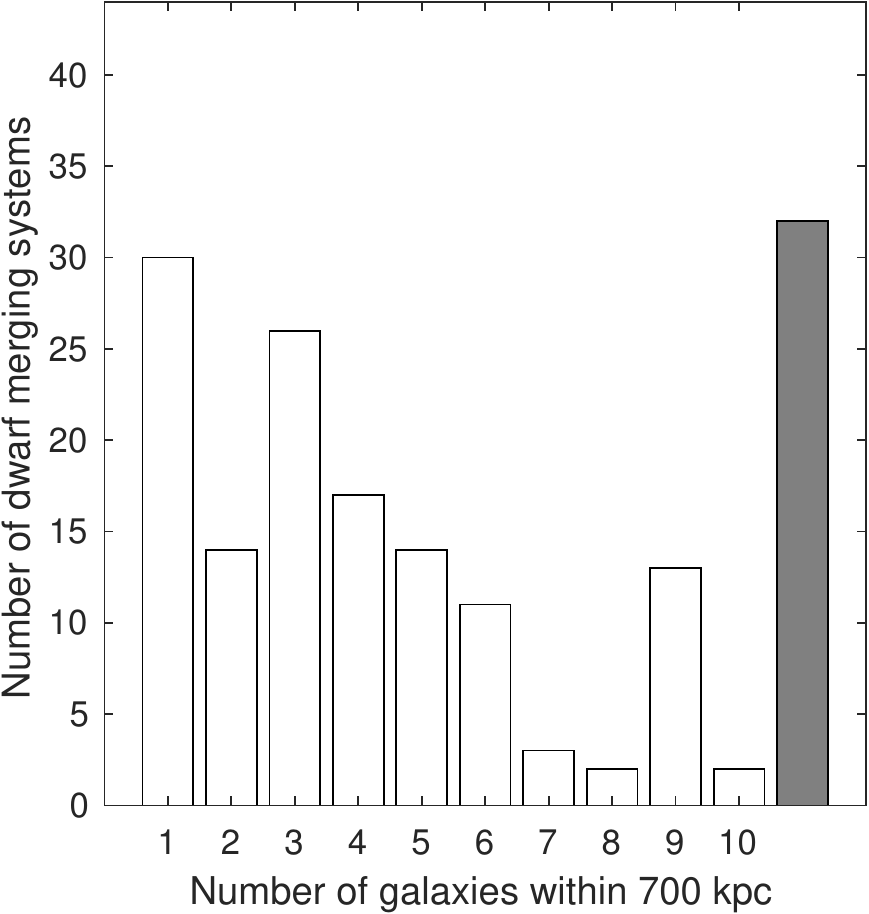}
\caption{Total number of galaxies, including both giants and dwarfs, within an area coverage of 700 kpc radius and $\pm$700 km/s line of sight radial velocity around merging dwarf systems. The last gray bar represents number of merging dwarf systems which have more than 10 neighbor galaxies. }
\label{denshist}
\end{figure}

\begin{figure}
\includegraphics[width = 8cm]{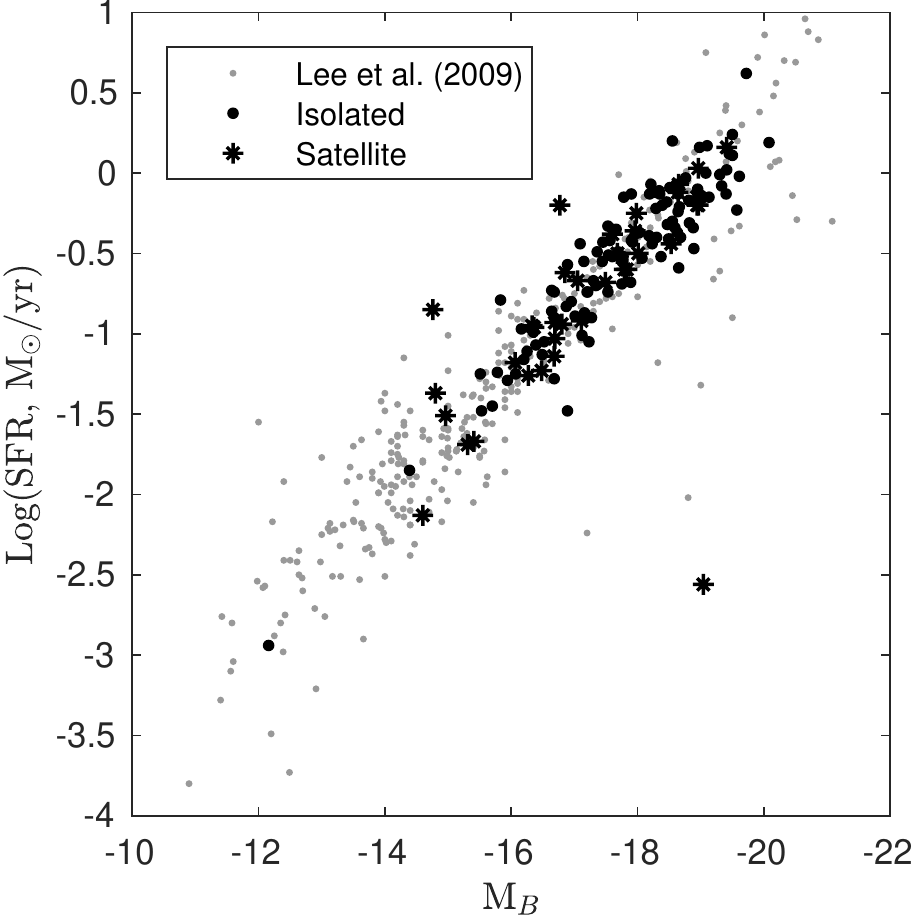}
\caption{Comparison of the star-formation rates of satellite (star)  and isolated (dot) merging dwarfs systems. We also show the local volume ($<$11 Mpc) star-forming galaxy sample of \cite{Lee09} in gray. }
\label{setsfr}
\end{figure}

We now turn to the surrounding environment of our merging dwarf systems. For this work we characterize the surrounding environment by searching for neighboring giant galaxies (M$_{K}$ $<$ -20 mag, corresponding to a stellar mass of $>$10$^{10}$), within a sky projected distance of less than 700 kpc, and a relative line of sight radial velocity of less than $\pm$700 km/s. This is the similar criteria that we have previously used to search for isolated early-type dwarf galaxies in \cite{Paudel14}.

We find that only 41 dwarf galaxy merging systems have giant neighbors. he median stellar mass of the giant neighbors is 6$\times$10$^{10}$ M$_{\sun}$. For convenience, we called them satellite merging dwarf system and the rests are isolated merging dwarf systems, here after. Among 41 satellite merging systems, there are 19 `I' class systems (interacting dwarf pairs) where we identify ongoing interaction between dwarf galaxies. Shell features are found in 10 systems and the remaining 12 are a mixture of E/T/S classes. 

Interestingly, all three early-type merging dwarfs are located at large sky projected distance from the giant galaxies, beyond 700 kpc. In fact, \cite{Paudel14} already pointed out that Id10080227 is a compact elliptical galaxy (cE \cite{Chilingarian09}), located in isolation, which may have formed through the merging of dwarf galaxies.

In Figure \ref{setprop}, we show a phase space diagram of the satellite merging dwarf systems. It is clear from this figure that our satellite merging dwarf systems are located comparatively farther than the distance of the LMC-SMC system is from the Milky-Way (MW). We also highlight the position of UGC 4703, which we studied as an LMC-SMC-MW analog in \cite{Paudel17}, and lies in a similar region in phase space. The dashed line represents the escape velocity as a function of radius for a Milky-Way like galaxy, based on the best match model to the Milky-way from \cite{Klypin02}. The two highlighted interacting dwarf pairs, LMC-SMC and  UGC 4703 are located at small radius and large velocity, near the escape velocity boundary, perhaps indicating they are recent infallers into their host \citep{Rhee17}. It seems that only half of the satellite merging dwarf systems are clearly bound to their hosts, (assuming their hosts are MW-like), i.e located below the escape velocity line. The rest are scattered well beyond the escape velocity boundary, and often at distances $>$400 kpc which is at least twice the Virial radius of a MW-like galaxy. Thus it is probable that many of these are not bound to their hosts, and in many cases our selection criteria of 700 kpc search radius and $\pm$700 km/s velocity range is not robust enough to characterize if our merging dwarfs are hosted by a nearest giant host. The phase space diagram also reveals that there is no special difference between satellite interacting dwarf pairs (shown as empty circle symbols) and the rest of the sample, that have likely already merged (shown as black dot), in their location of their phase-space diagram. 
  
We compare the star-formation rates of candidate satellites and isolated merging dwarf system, see Figure \ref{setsfr}. The black dot represent the satellite candidates and blue dots represent isolated. From this figure, it is clear that both the isolated and satellite merging dwarf system have similar star formation properties compared with \cite{Lee09}. We also find only a marginal difference in the distribution of gas mass fraction of satellite and isolated dwarf systems with the median values  1.04 and 1.09, respectively. This is slightly contradictory to the finding of S15, where they found interacting dwarfs located near to the giant galaxy are likely to have a lower gas mass fraction.

We also attempt using number density to characterize the surrounding environment of merging dwarf systems. For this we, simply searched the number of galaxies, both giant and dwarf, within the above mentioned search area (i.e within 700 kpc radius and $\pm$700 km/s line of sight radial velocity). For this, we also remove those merging dwarf systems which have a line of sight radial velocity less than 900 km/s to avoid distance uncertainties of nearby galaxies.

We find that a significant fraction, 30 out 177, merging dwarf systems have no neighbor, not even another dwarf galaxy, within our search area. In contrast,  more than 10 neighbors are found only for 32 cases, and they are mostly interacting satellites. We show a simple histogram of the number of galaxies (which include both giants and dwarfs) found in the search area in Figure \ref{denshist}. The last bin (the gray histogram) represents the number of merging dwarf systems which have more than 10 neighbors within our search area. From this figure, it is clear that the probability of finding a merging dwarf system increases in low density environments. The median neighbor number of merging dwarf systems in this sample is 4.

\section{Conclusions and Remarks}\label{cons}

We have collected a catalog of 177 merging dwarf systems, spanning the stellar mass range from 10$^{7}$ to 10$^{10}$ M$_{\sun}$ in a redshift range z $<$ 0.02.
The sample is overwhelmingly dominated by star-forming galaxies and they are located significantly below the red-sequence in the observed color-magnitude relation. The fraction of early-type dwarf galaxies is only 3 out of 177. Star forming objects may be preferentially selected because of the criterion to have a redshift and it is easy to measure redshift from emission line of star-forming galaxies than absorption line of non star-forming galaxies.

We classify the morphology of the low surface brightness feature into various categories such as shells, stellar streams, loops, Antennae-like systems, or simply interacting. These different types of low surface brightness features may hint at objects in different stages of their interactions. For example, the shell feature might be the product of a complete coalescence, while two well separated interacting dwarfs are probably in the earlier stages of their interaction. There are three dwarf galaxies (Id0202-0922, Id1448-0342, Id14503534) that can be considered dwarf analogues to the Antennae-system (NGC 4038/4039). 

A potential problem with these types of catalog is that they are inherently inhomogeneous and incomplete. Because they are selected from visual inspection of low-surface brightness feature, this depends on the depth of the imaging survey, and on how well defined the tidal features are. As a result, this is in many ways very subjective. We certainly caution on the completeness of the catalog and there maybe many possible biases in our selection procedure. For example, dwarf galaxies with tidal features whose origin is unclear and are located near to a giant (M$_{*}$ $>$ 10$^{10}$ ) host galaxy have been selectively removed. That may lead to an artificial reduction in the number of merging dwarf systems near giant galaxies. 

However, more isolated dwarf interacting pairs do not suffer this issue, as there is no uncertainty as to whether a giant galaxy is responsible for the observed fine structure (e.g. tidal streams, tails, shells, etc). Therefore, we believe our sample will be more complete for these kinds of objects, as long as the interacting pairs show similar low surface brightness features as presented by our sample. We believe that it makes physical sense that dwarf systems struggle to merge in the presence of a nearby giant galaxy. Dwarf galaxies have small escape velocities, owing to their small masses. As a result only a small amount of peculiar motion, due to the potential well of a giant galaxy, might be enough to make it nearly impossible for dwarfs to meet at low enough velocities to merge. Thus, we suspect that our selection criteria maybe simply enhancing a real dependency on distance to the nearest giant galaxy. In any case, we find that there is no significant difference in the phase-space diagram of dwarf interacting pair (I class) and the rest of the sample.

In conclusion, we present a large set of interacting and merging dwarf systems, including aperture photometry in UV and optical bands, as well as stellar masses, star formation rates, gas masses and stellar mass ratios. This data might be useful for detailed studies of dwarf-dwarf interactions in the near future.

\acknowledgements
P.S. acknowledges the support by Samsung Science \& Technology Foundation under Project Number SSTF-BA1501-0. S.-J.Y. acknowledges support from the Center for Galaxy Evolution Research (No. 2010-0027910) through the NRF of Korea and from the Yonsei University Observatory -- KASI Joint Research Program (2018). P.C-C. was supported by CONICYT (Chile) through Programa Nacional de Becas de Doctorado 2014 folio 21140882.

This study is based on the archival images and spectra from the Sloan Digital Sky Survey and Legacy Survey Data. Their full acknowledgment can be found at http://www.sdss.org/collaboration/credits.html and  http://legacysurvey.org/acknowledgment/, respectively. Funding for the SDSS has been provided by the Alfred P. Sloan Foundation, the Participating Institutions, the National Science Foundation, the U.S. Department of Energy, the National Aeronautics and Space Administration, the Japanese Monbukagakusho, the Max Planck Society, and the Higher Education Funding Council for England. The SDSS Web Site is http://www.sdss.org/. The Legacy Surveys imaging of the DESI footprint is supported by the Director, Office of Science, Office of High Energy Physics of the U.S. Department of Energy under Contract No. DE-AC02-05CH1123, by the National Energy Research Scientific Computing Center, a DOE Office of Science User Facility under the same contract; and by the U.S. National Science Foundation, Division of Astronomical Sciences under Contract No. AST-0950945 to NOAO. We also made use of the GALEX all-sky survey imaging data. The GALEX is operated for NASA by the California Institute of Technology under NASA contract NAS5-98034. We also acknowledge the use of NASA's Astrophysics Data System Bibliographic Services and the NASA/IPAC Extragalactic Database (NED). We also made use of archival data from Canada-France-Hawaii Telescope (CFHT) which is operated by the National Research Council (NRC) of Canada, the Institute National des Sciences de l\'Univers of the Centre National de la Recherche Scientifique of France and the University of Hawaii.

\newpage
 \appendix
\section{Notes on selected individual system}
In this Section, we provide a short list of previously published studies on individual objects in our sample. We npte that the list is not complete or fully comprehensive, but we hope it provides a useful starting point for readers with an interest in a specific object or merging system.\\
\\
\textbf{Id01130052}: A gas rich low metallicity dwarf galaxy, \cite{Ekta08} find disturbed HI velocity field and suggest ongoing merger.\\
\textbf{Id0202-0922}: Dwarf antennae system produced by merging two gas-rich dwarf galaxies. A detail study of of the system from HI data has been submitted for the publication, Paudel et al.\\
\textbf{Id02032202}:  This galaxy is located in isolation and \cite{Sengupta12} reported ongoing minor merger in this galaxy. They detected an a symmetry feature in HI map.\\
\textbf{Id0851-0221}:  ARP 257: from catalog of interacting galaxies \citep{Arp66}\\
\textbf{Id08580619}: Interacting dwarf pair in the vicinity of an isolated spiral galaxy NGC 2718. \cite{Paudel17b} reported the system as LMC-SMC-MW analoge. \\
\textbf{Id09003543}: Arp 202: from catalog of interacting galaxies. A detail study of the system has been performed in \cite{Sengupta14} and they reported formation of tidal dwarf galaxies of stellar mass 2$\times$10$^{8}$ M$_{\sun}$. \\
\textbf{Id09002536}: An isolated galaxy. \cite{Chengalur15} identified a disturbed HI morphology and argued that the galaxy has suffered recent minor merger.\\
\textbf{Id09562849}: A merging dwarf candidate \cite{Annibali16}\\
\textbf{Id10080227}: A merger origin compact early-type galaxy \citep{Paudel14}\\
\textbf{Id10545418}: Interacting pair studied in Local Volume TiNy Titan \citep{Pearson16}\\
\textbf{ Id11451711}: Interacting dwarf galaxies in the outskirt of a group environment \citep{Paudel13}.\\
\textbf{Id1148-0138}: \cite{Lelli14} studied this galaxy and concluded that star-formation rate is enhanced due to merger/interaction in recent past.\\
\textbf{Id12250548}: VCC848,  a merging Blue Compact Dwarf in Virgo cluster.\\
\textbf{ Id12284405}: NGC 4449: Interacting dwarf galaxies reported by  \citep{Delgado12,Rich12}\\
\textbf{Id12304138}: Interacting pair studied in Local Volume TiNy Titan \citep{Pearson16}\\
\textbf{Id12474709}:  ARP 277\\
\textbf{Id14503534}: Interacting pair studied in Local Volume TiNy Titan \citep{Pearson16}\\
\textbf{Id14503534}: Part of  TiNy Titan, dwarf interacting pair studied in \cite{Privon17}.\\

\newpage
\section{Figure catalog}
\begin{figure*}[h]
\centering
\includegraphics[width=17cm, page=1]{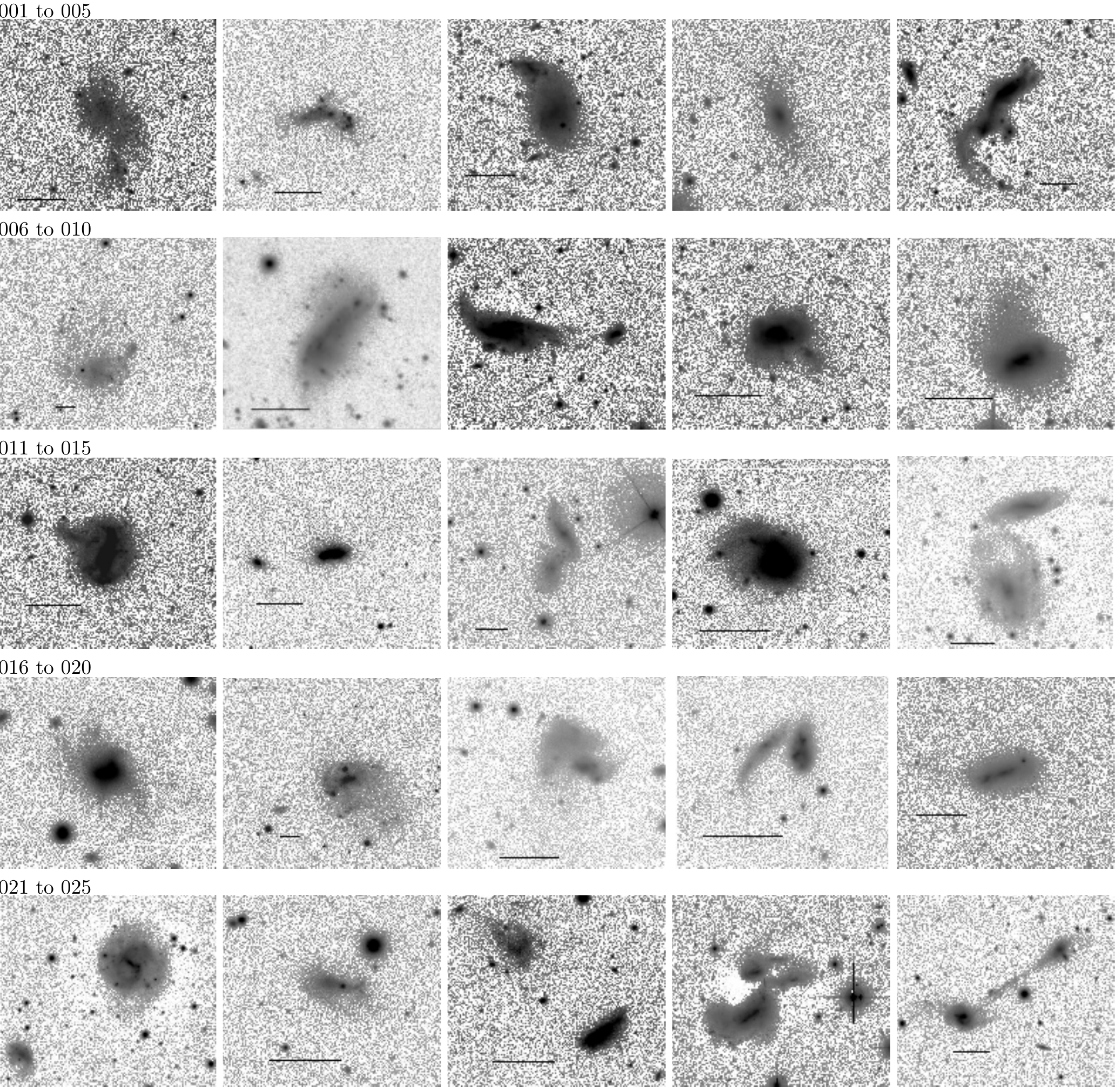}
\caption{These postage images are prepared from fits images downloaded from various archive. On the top of each row, we list identification of these galaxies according to Table \ref{mtb}. The field of view and color stretching is arbitrarily chosen to make best view of both interacting galaxies and low-surface brightness features.  An image scale of 30" is shown by the black horizontal bar.}
\label{fcat}
\end{figure*}
\newpage
\centering
\includegraphics[width=17cm, page=2]{Allfig_n.pdf}
Continue Figure \ref{fcat}
\newpage
\centering
\includegraphics[width=17cm, page=3]{Allfig_n.pdf}
Continue Figure \ref{fcat}
\newpage
\centering
\includegraphics[width=17cm, page=4]{Allfig_n.pdf}
Continue Figure \ref{fcat}
\newpage
\centering
\includegraphics[width=17cm, page=5]{Allfig_n.pdf}
Continue Figure \ref{fcat}
\newpage
\centering
\includegraphics[width=17cm, page=6]{Allfig_n.pdf}
Continue Figure \ref{fcat}


\newpage
\begin{thebibliography}{}
\expandafter\ifx\csname natexlab\endcsname\relax\def\natexlab#1{#1}\fi
\providecommand{\url}[1]{\href{#1}{#1}}
\providecommand{\dodoi}[1]{doi:~\href{http://doi.org/#1}{\nolinkurl{#1}}}
\providecommand{\doeprint}[1]{\href{http://ascl.net/#1}{\nolinkurl{http://ascl.net/#1}}}
\providecommand{\doarXiv}[1]{\href{https://arxiv.org/abs/#1}{\nolinkurl{https://arxiv.org/abs/#1}}}

\bibitem[{{Abazajian} {et~al.}(2009){Abazajian}, {Adelman-McCarthy},
  {Ag{\"u}eros}, {Allam}, {Allende Prieto}, {An}, {Anderson}, {Anderson},
  {Annis}, {Bahcall}, \& et~al.}]{Abazajian09}
{Abazajian}, K.~N., {Adelman-McCarthy}, J.~K., {Ag{\"u}eros}, M.~A., {et~al.}
  2009, \apjs, 182, 543, \dodoi{10.1088/0067-0049/182/2/543}

\bibitem[{{Abraham} \& {van Dokkum}(2014)}]{Abraham14}
{Abraham}, R.~G., \& {van Dokkum}, P.~G. 2014, \pasp, 126, 55,
  \dodoi{10.1086/674875}

\bibitem[{{Aihara} {et~al.}(2011){Aihara}, {Allende Prieto}, {An}, {Anderson},
  {Aubourg}, {Balbinot}, {Beers}, {Berlind}, {Bickerton}, {Bizyaev}, {Blanton},
  {Bochanski}, {Bolton}, {Bovy}, {Brandt}, {Brinkmann}, {Brown}, {Brownstein},
  {Busca}, {Campbell}, {Carr}, {Chen}, {Chiappini}, {Comparat}, {Connolly},
  {Cortes}, {Croft}, {Cuesta}, {da Costa}, {Davenport}, {Dawson}, {Dhital},
  {Ealet}, {Ebelke}, {Edmondson}, {Eisenstein}, {Escoffier}, {Esposito},
  {Evans}, {Fan}, {Femen{\'{\i}}a Castell{\'a}}, {Font-Ribera}, {Frinchaboy},
  {Ge}, {Gillespie}, {Gilmore}, {Gonz{\'a}lez Hern{\'a}ndez}, {Gott}, {Gould},
  {Grebel}, {Gunn}, {Hamilton}, {Harding}, {Harris}, {Hawley}, {Hearty}, {Ho},
  {Hogg}, {Holtzman}, {Honscheid}, {Inada}, {Ivans}, {Jiang}, {Johnson},
  {Jordan}, {Jordan}, {Kazin}, {Kirkby}, {Klaene}, {Knapp}, {Kneib},
  {Kochanek}, {Koesterke}, {Kollmeier}, {Kron}, {Lampeitl}, {Lang}, {Le Goff},
  {Lee}, {Lin}, {Long}, {Loomis}, {Lucatello}, {Lundgren}, {Lupton}, {Ma},
  {MacDonald}, {Mahadevan}, {Maia}, {Makler}, {Malanushenko}, {Malanushenko},
  {Mandelbaum}, {Maraston}, {Margala}, {Masters}, {McBride}, {McGehee},
  {McGreer}, {M{\'e}nard}, {Miralda-Escud{\'e}}, {Morrison}, {Mullally},
  {Muna}, {Munn}, {Murayama}, {Myers}, {Naugle}, {Neto}, {Nguyen}, {Nichol},
  {O'Connell}, {Ogando}, {Olmstead}, {Oravetz}, {Padmanabhan},
  {Palanque-Delabrouille}, {Pan}, {Pandey}, {P{\^a}ris}, {Percival},
  {Petitjean}, {Pfaffenberger}, {Pforr}, {Phleps}, {Pichon}, {Pieri}, {Prada},
  {Price-Whelan}, {Raddick}, {Ramos}, {Reyl{\'e}}, {Rich}, {Richards}, {Rix},
  {Robin}, {Rocha-Pinto}, {Rockosi}, {Roe}, {Rollinde}, {Ross}, {Ross},
  {Rossetto}, {S{\'a}nchez}, {Sayres}, {Schlegel}, {Schlesinger}, {Schmidt},
  {Schneider}, {Sheldon}, {Shu}, {Simmerer}, {Simmons}, {Sivarani}, {Snedden},
  {Sobeck}, {Steinmetz}, {Strauss}, {Szalay}, {Tanaka}, {Thakar}, {Thomas},
  {Tinker}, {Tofflemire}, {Tojeiro}, {Tremonti}, {Vandenberg}, {Vargas
  Maga{\~n}a}, {Verde}, {Vogt}, {Wake}, {Wang}, {Weaver}, {Weinberg}, {White},
  {White}, {Yanny}, {Yasuda}, {Yeche}, \& {Zehavi}}]{Aihara11}
{Aihara}, H., {Allende Prieto}, C., {An}, D., {et~al.} 2011, \apjs, 193, 29,
  \dodoi{10.1088/0067-0049/193/2/29}

\bibitem[{{Amorisco} {et~al.}(2014){Amorisco}, {Evans}, \& {van de
  Ven}}]{Amorisco14}
{Amorisco}, N.~C., {Evans}, N.~W., \& {van de Ven}, G. 2014, \nat, 507, 335,
  \dodoi{10.1038/nature12995}

\bibitem[{{Annibali} {et~al.}(2016){Annibali}, {Nipoti}, {Ciotti}, {Tosi},
  {Aloisi}, {Bellazzini}, {Cignoni}, {Cusano}, {Paris}, \&
  {Sacchi}}]{Annibali16}
{Annibali}, F., {Nipoti}, C., {Ciotti}, L., {et~al.} 2016, \apjl, 826, L27,
  \dodoi{10.3847/2041-8205/826/2/L27}

\bibitem[{{Arp}(1966)}]{Arp66}
{Arp}, H. 1966, \apjs, 14, 1, \dodoi{10.1086/190147}

\bibitem[{{Barnes} \& {Hibbard}(2009)}]{Barnes09}
{Barnes}, J.~E., \& {Hibbard}, J.~E. 2009, \aj, 137, 3071,
  \dodoi{10.1088/0004-6256/137/2/3071}

\bibitem[{{Bell} {et~al.}(2003){Bell}, {McIntosh}, {Katz}, \&
  {Weinberg}}]{Bell03}
{Bell}, E.~F., {McIntosh}, D.~H., {Katz}, N., \& {Weinberg}, M.~D. 2003, \apjs,
  149, 289, \dodoi{10.1086/378847}

\bibitem[{{Besla} {et~al.}(2016){Besla}, {Mart{\'{\i}}nez-Delgado}, {van der
  Marel}, {Beletsky}, {Seibert}, {Schlafly}, {Grebel}, \& {Neyer}}]{Besla16}
{Besla}, G., {Mart{\'{\i}}nez-Delgado}, D., {van der Marel}, R.~P., {et~al.}
  2016, \apj, 825, 20, \dodoi{10.3847/0004-637X/825/1/20}

\bibitem[{{Blum} {et~al.}(2016){Blum}, {Burleigh}, {Dey}, {Schlegel},
  {Meisner}, {Levi}, {Myers}, {Lang}, {Moustakas}, {Patej}, {Valdes}, {Kneib},
  {Huanyuan}, {Nord}, {Olsen}, {Delubac}, {Saha}, {James}, {Walker}, \& {DECaLS
  Team}}]{Blum16}
{Blum}, R.~D., {Burleigh}, K., {Dey}, A., {et~al.} 2016, in American
  Astronomical Society Meeting Abstracts, Vol. 228, American Astronomical
  Society Meeting Abstracts \#228, 317.01

\bibitem[{{Boselli} \& {Gavazzi}(2006)}]{Boselli06}
{Boselli}, A., \& {Gavazzi}, G. 2006, \pasp, 118, 517, \dodoi{10.1086/500691}

\bibitem[{{Brinchmann} {et~al.}(2004){Brinchmann}, {Charlot}, {White},
  {Tremonti}, {Kauffmann}, {Heckman}, \& {Brinkmann}}]{Brinchmann04}
{Brinchmann}, J., {Charlot}, S., {White}, S.~D.~M., {et~al.} 2004, \mnras, 351,
  1151, \dodoi{10.1111/j.1365-2966.2004.07881.x}

\bibitem[{{Chengalur} {et~al.}(2015){Chengalur}, {Pustilnik}, {Makarov},
  {Perepelitsyna}, {Safonova}, \& {Karachentsev}}]{Chengalur15}
{Chengalur}, J.~N., {Pustilnik}, S.~A., {Makarov}, D.~I., {et~al.} 2015,
  \mnras, 448, 1634, \dodoi{10.1093/mnras/stv086}

\bibitem[{{Chilingarian}(2009)}]{Chilingarian09}
{Chilingarian}, I.~V. 2009, \mnras, 394, 1229,
  \dodoi{10.1111/j.1365-2966.2009.14450.x}

\bibitem[{{Coleman} {et~al.}(2004){Coleman}, {Da Costa}, {Bland-Hawthorn},
  {Mart{\'{\i}}nez-Delgado}, {Freeman}, \& {Malin}}]{Coleman04}
{Coleman}, M., {Da Costa}, G.~S., {Bland-Hawthorn}, J., {et~al.} 2004, \aj,
  127, 832, \dodoi{10.1086/381298}

\bibitem[{{Conselice} \& {Gallagher}(1999)}]{Conselice99}
{Conselice}, C.~J., \& {Gallagher}, III, J.~S. 1999, \aj, 117, 75,
  \dodoi{10.1086/300697}

\bibitem[{{Conselice} {et~al.}(2009){Conselice}, {Yang}, \&
  {Bluck}}]{Conselice09}
{Conselice}, C.~J., {Yang}, C., \& {Bluck}, A.~F.~L. 2009, \mnras, 394, 1956,
  \dodoi{10.1111/j.1365-2966.2009.14396.x}

\bibitem[{{Courtois} \& {Tully}(2015)}]{Courtois15}
{Courtois}, H.~M., \& {Tully}, R.~B. 2015, \mnras, 447, 1531,
  \dodoi{10.1093/mnras/stu2405}

\bibitem[{{Crnojevi{\'c}} {et~al.}(2014){Crnojevi{\'c}}, {Sand}, {Caldwell},
  {Guhathakurta}, {McLeod}, {Seth}, {Simon}, {Strader}, \&
  {Toloba}}]{Crnojevic14}
{Crnojevi{\'c}}, D., {Sand}, D.~J., {Caldwell}, N., {et~al.} 2014, ArXiv
  e-prints.
\newblock \doarXiv{1409.4776}

\bibitem[{{Duc} {et~al.}(2014{\natexlab{a}}){Duc}, {Paudel}, {McDermid},
  {Cuillandre}, {Serra}, {Bournaud}, {Cappellari}, \& {Emsellem}}]{Duc14}
{Duc}, P.-A., {Paudel}, S., {McDermid}, R.~M., {et~al.} 2014{\natexlab{a}},
  \mnras, \dodoi{10.1093/mnras/stu330}

\bibitem[{{Duc} \& {Renaud}(2013)}]{Duc13}
{Duc}, P.-A., \& {Renaud}, F. 2013, in Lecture Notes in Physics, Berlin
  Springer Verlag, Vol. 861, Lecture Notes in Physics, Berlin Springer Verlag,
  ed. J.~{Souchay}, S.~{Mathis}, \& T.~{Tokieda}, 327

\bibitem[{{Duc} {et~al.}(2011){Duc}, {Cuillandre}, {Serra}, {Michel-Dansac},
  {Ferriere}, {Alatalo}, {Blitz}, {Bois}, {Bournaud}, {Bureau}, {Cappellari},
  {Davies}, {Davis}, {de Zeeuw}, {Emsellem}, {Khochfar}, {Krajnovi{\'c}},
  {Kuntschner}, {Lablanche}, {McDermid}, {Morganti}, {Naab}, {Oosterloo},
  {Sarzi}, {Scott}, {Weijmans}, \& {Young}}]{Duc11}
{Duc}, P.-A., {Cuillandre}, J.-C., {Serra}, P., {et~al.} 2011, \mnras, 417,
  863, \dodoi{10.1111/j.1365-2966.2011.19137.x}

\bibitem[{{Duc} {et~al.}(2014{\natexlab{b}}){Duc}, {Cuillandre}, {Karabal},
  {Cappellari}, {Alatalo}, {Blitz}, {Bournaud}, {Bureau}, {Crocker}, {Davies},
  {Davis}, {de Zeeuw}, {Emsellem}, {Khochfar}, {Krajnovic}, {Kuntschner},
  {McDermid}, {Michel-Dansac}, {Morganti}, {Naab}, {Oosterloo}, {Paudel},
  {Sarzi}, {Scott}, {Serra}, {Weijmans}, \& {Young}}]{Duc14b}
{Duc}, P.-A., {Cuillandre}, J.-C., {Karabal}, E., {et~al.} 2014{\natexlab{b}},
  ArXiv e-prints.
\newblock \doarXiv{1410.0981}

\bibitem[{{Duc} {et~al.}(2015){Duc}, {Cuillandre}, {Karabal}, {Cappellari},
  {Alatalo}, {Blitz}, {Bournaud}, {Bureau}, {Crocker}, {Davies}, {Davis}, {de
  Zeeuw}, {Emsellem}, {Khochfar}, {Krajnovi{\'c}}, {Kuntschner}, {McDermid},
  {Michel-Dansac}, {Morganti}, {Naab}, {Oosterloo}, {Paudel}, {Sarzi}, {Scott},
  {Serra}, {Weijmans}, \& {Young}}]{Duc15}
---. 2015, \mnras, 446, 120, \dodoi{10.1093/mnras/stu2019}

\bibitem[{{Ekta} {et~al.}(2008){Ekta}, {Chengalur}, \& {Pustilnik}}]{Ekta08}
{Ekta}, {Chengalur}, J.~N., \& {Pustilnik}, S.~A. 2008, \mnras, 391, 881,
  \dodoi{10.1111/j.1365-2966.2008.13928.x}

\bibitem[{{Eneev} {et~al.}(1973){Eneev}, {Kozlov}, \& {Sunyaev}}]{Eneev73}
{Eneev}, T.~M., {Kozlov}, N.~N., \& {Sunyaev}, R.~A. 1973, \aap, 22, 41

\bibitem[{{Gallagher} {et~al.}(1984){Gallagher}, {Hunter}, \&
  {Tutukov}}]{Gallagher84}
{Gallagher}, III, J.~S., {Hunter}, D.~A., \& {Tutukov}, A.~V. 1984, \apj, 284,
  544, \dodoi{10.1086/162437}

\bibitem[{{Gallagher} \& {Parker}(2010)}]{Gallagher10}
{Gallagher}, III, J.~S., \& {Parker}, A. 2010, \apj, 722, 1962,
  \dodoi{10.1088/0004-637X/722/2/1962}

\bibitem[{{Geha} {et~al.}(2012){Geha}, {Blanton}, {Yan}, \& {Tinker}}]{Geha12}
{Geha}, M., {Blanton}, M.~R., {Yan}, R., \& {Tinker}, J.~L. 2012, \apj, 757,
  85, \dodoi{10.1088/0004-637X/757/1/85}

\bibitem[{{Geha} {et~al.}(2005){Geha}, {Guhathakurta}, \& {van der
  Marel}}]{Geha05}
{Geha}, M., {Guhathakurta}, P., \& {van der Marel}, R.~P. 2005, \aj, 129, 2617,
  \dodoi{10.1086/430188}

\bibitem[{{Gil de Paz} {et~al.}(2003){Gil de Paz}, {Madore}, \&
  {Pevunova}}]{Gil03}
{Gil de Paz}, A., {Madore}, B.~F., \& {Pevunova}, O. 2003, \apjs, 147, 29,
  \dodoi{10.1086/374737}

\bibitem[{{Giovanelli} {et~al.}(2005){Giovanelli}, {Haynes}, {Kent},
  {Perillat}, {Saintonge}, {Brosch}, {Catinella}, {Hoffman}, {Stierwalt},
  {Spekkens}, {Lerner}, {Masters}, {Momjian}, {Rosenberg}, {Springob},
  {Boselli}, {Charmandaris}, {Darling}, {Davies}, {Garcia Lambas}, {Gavazzi},
  {Giovanardi}, {Hardy}, {Hunt}, {Iovino}, {Karachentsev}, {Karachentseva},
  {Koopmann}, {Marinoni}, {Minchin}, {Muller}, {Putman}, {Pantoja}, {Salzer},
  {Scodeggio}, {Skillman}, {Solanes}, {Valotto}, {van Driel}, \& {van
  Zee}}]{Giovanelli05}
{Giovanelli}, R., {Haynes}, M.~P., {Kent}, B.~R., {et~al.} 2005, \aj, 130,
  2598, \dodoi{10.1086/497431}

\bibitem[{{Graham} {et~al.}(2012){Graham}, {Spitler}, {Forbes}, {Lisker},
  {Moore}, \& {Janz}}]{Graham12}
{Graham}, A.~W., {Spitler}, L.~R., {Forbes}, D.~A., {et~al.} 2012, \apj, 750,
  121, \dodoi{10.1088/0004-637X/750/2/121}

\bibitem[{{Gwyn}(2008)}]{Gwyn08}
{Gwyn}, S.~D.~J. 2008, \pasp, 120, 212, \dodoi{10.1086/526794}

\bibitem[{{James} {et~al.}(2013){James}, {Tsamis}, {Barlow}, {Walsh}, \&
  {Westmoquette}}]{James13}
{James}, B.~L., {Tsamis}, Y.~G., {Barlow}, M.~J., {Walsh}, J.~R., \&
  {Westmoquette}, M.~S. 2013, \mnras, 428, 86, \dodoi{10.1093/mnras/sts004}

\bibitem[{{Janz} \& {Lisker}(2009)}]{Janz09}
{Janz}, J., \& {Lisker}, T. 2009, \apjl, 696, L102,
  \dodoi{10.1088/0004-637X/696/1/L102}

\bibitem[{{Johnson}(2013)}]{Johnson13}
{Johnson}, M. 2013, \aj, 145, 146, \dodoi{10.1088/0004-6256/145/6/146}

\bibitem[{{Kennicutt}(1998)}]{Kennicutt98}
{Kennicutt}, Jr., R.~C. 1998, \araa, 36, 189,
  \dodoi{10.1146/annurev.astro.36.1.189}

\bibitem[{{Kim} {et~al.}(2012){Kim}, {Sheth}, {Hinz}, {Lee}, {Zaritsky},
  {Gadotti}, {Knapen}, {Schinnerer}, {Ho}, {Laurikainen}, {Salo},
  {Athanassoula}, {Bosma}, {de Swardt}, {Mu{\~n}oz-Mateos}, {Madore},
  {Comer{\'o}n}, {Regan}, {Men{\'e}ndez-Delmestre}, {Gil de Paz}, {Seibert},
  {Laine}, {Erroz-Ferrer}, \& {Mizusawa}}]{Kim12}
{Kim}, T., {Sheth}, K., {Hinz}, J.~L., {et~al.} 2012, \apj, 753, 43,
  \dodoi{10.1088/0004-637X/753/1/43}

\bibitem[{{Klypin} {et~al.}(2002){Klypin}, {Zhao}, \& {Somerville}}]{Klypin02}
{Klypin}, A., {Zhao}, H., \& {Somerville}, R.~S. 2002, \apj, 573, 597,
  \dodoi{10.1086/340656}

\bibitem[{{Kormendy} {et~al.}(2009){Kormendy}, {Fisher}, {Cornell}, \&
  {Bender}}]{Kormendy09}
{Kormendy}, J., {Fisher}, D.~B., {Cornell}, M.~E., \& {Bender}, R. 2009, \apjs,
  182, 216, \dodoi{10.1088/0067-0049/182/1/216}

\bibitem[{{Lee} {et~al.}(2009){Lee}, {Gil de Paz}, {Tremonti}, {Kennicutt},
  {Salim}, {Bothwell}, {Calzetti}, {Dalcanton}, {Dale}, {Engelbracht}, {Funes},
  {Johnson}, {Sakai}, {Skillman}, {van Zee}, {Walter}, \& {Weisz}}]{Lee09}
{Lee}, J.~C., {Gil de Paz}, A., {Tremonti}, C., {et~al.} 2009, \apj, 706, 599,
  \dodoi{10.1088/0004-637X/706/1/599}

\bibitem[{{Lelli} {et~al.}(2014){Lelli}, {Verheijen}, \&
  {Fraternali}}]{Lelli14}
{Lelli}, F., {Verheijen}, M., \& {Fraternali}, F. 2014, \mnras, 445, 1694,
  \dodoi{10.1093/mnras/stu1804}

\bibitem[{{Leroy} {et~al.}(2008){Leroy}, {Walter}, {Brinks}, {Bigiel}, {de
  Blok}, {Madore}, \& {Thornley}}]{Leroy08}
{Leroy}, A.~K., {Walter}, F., {Brinks}, E., {et~al.} 2008, \aj, 136, 2782,
  \dodoi{10.1088/0004-6256/136/6/2782}

\bibitem[{{Lisker}(2009)}]{Lisker09}
{Lisker}, T. 2009, Astronomische Nachrichten, 330, 1043,
  \dodoi{10.1002/asna.200911291}

\bibitem[{{Martin} {et~al.}(2005){Martin}, {Fanson}, {Schiminovich},
  {Morrissey}, {Friedman}, {Barlow}, {Conrow}, {Grange}, {Jelinsky},
  {Milliard}, {Siegmund}, {Bianchi}, \& {Byun}}]{Martin05}
{Martin}, D.~C., {Fanson}, J., {Schiminovich}, D., {et~al.} 2005, \apjl, 619,
  L1, \dodoi{10.1086/426387}

\bibitem[{{Mart{\'{\i}}nez-Delgado} {et~al.}(2012){Mart{\'{\i}}nez-Delgado},
  {Romanowsky}, {Gabany}, {Annibali}, {Arnold}, {Fliri}, {Zibetti}, {van der
  Marel}, {Rix}, \& {Chonis}}]{Delgado12}
{Mart{\'{\i}}nez-Delgado}, D., {Romanowsky}, A.~J., {Gabany}, R.~J., {et~al.}
  2012, \apjl, 748, L24, \dodoi{10.1088/2041-8205/748/2/L24}

\bibitem[{{Meyer} {et~al.}(2014){Meyer}, {Lisker}, {Janz}, \&
  {Papaderos}}]{Meyer14}
{Meyer}, H.~T., {Lisker}, T., {Janz}, J., \& {Papaderos}, P. 2014, \aap, 562,
  A49, \dodoi{10.1051/0004-6361/201220700}

\bibitem[{{Meyer} {et~al.}(2004){Meyer}, {Zwaan}, {Webster}, {Staveley-Smith},
  {Ryan-Weber}, {Drinkwater}, {Barnes}, {Howlett}, {Kilborn}, {Stevens},
  {Waugh}, {Pierce}, {Bhathal}, {de Blok}, {Disney}, {Ekers}, {Freeman},
  {Garcia}, {Gibson}, {Harnett}, {Henning}, {Jerjen}, {Kesteven}, {Knezek},
  {Koribalski}, {Mader}, {Marquarding}, {Minchin}, {O'Brien}, {Oosterloo},
  {Price}, {Putman}, {Ryder}, {Sadler}, {Stewart}, {Stootman}, \&
  {Wright}}]{Meyer04}
{Meyer}, M.~J., {Zwaan}, M.~A., {Webster}, R.~L., {et~al.} 2004, \mnras, 350,
  1195, \dodoi{10.1111/j.1365-2966.2004.07710.x}

\bibitem[{{Mihos} {et~al.}(2017){Mihos}, {Harding}, {Feldmeier}, {Rudick},
  {Janowiecki}, {Morrison}, {Slater}, \& {Watkins}}]{Mihos17}
{Mihos}, J.~C., {Harding}, P., {Feldmeier}, J.~J., {et~al.} 2017, \apj, 834,
  16, \dodoi{10.3847/1538-4357/834/1/16}

\bibitem[{{Naab} {et~al.}(2007){Naab}, {Johansson}, {Ostriker}, \&
  {Efstathiou}}]{Naab07}
{Naab}, T., {Johansson}, P.~H., {Ostriker}, J.~P., \& {Efstathiou}, G. 2007,
  \apj, 658, 710, \dodoi{10.1086/510841}

\bibitem[{{Nidever} {et~al.}(2013){Nidever}, {Ashley}, {Slater}, {Ott},
  {Johnson}, {Bell}, {Stanimirovi{\'c}}, {Putman}, {Majewski}, {Simpson},
  {J{\"u}tte}, {Oosterloo}, \& {Butler Burton}}]{Nidever13}
{Nidever}, D.~L., {Ashley}, T., {Slater}, C.~T., {et~al.} 2013, \apjl, 779,
  L15, \dodoi{10.1088/2041-8205/779/2/L15}

\bibitem[{{Papaderos} {et~al.}(1996){Papaderos}, {Loose}, {Thuan}, \&
  {Fricke}}]{Papaderos96}
{Papaderos}, P., {Loose}, H.-H., {Thuan}, T.~X., \& {Fricke}, K.~J. 1996,
  \aaps, 120, 207

\bibitem[{{Patton} {et~al.}(2013){Patton}, {Torrey}, {Ellison}, {Mendel}, \&
  {Scudder}}]{Patton13}
{Patton}, D.~R., {Torrey}, P., {Ellison}, S.~L., {Mendel}, J.~T., \& {Scudder},
  J.~M. 2013, \mnras, 433, L59, \dodoi{10.1093/mnrasl/slt058}

\bibitem[{{Paturel} {et~al.}(2003){Paturel}, {Theureau}, {Bottinelli},
  {Gouguenheim}, {Coudreau-Durand}, {Hallet}, \& {Petit}}]{Paturel03}
{Paturel}, G., {Theureau}, G., {Bottinelli}, L., {et~al.} 2003, \aap, 412, 57,
  \dodoi{10.1051/0004-6361:20031412}

\bibitem[{{Paudel} {et~al.}(2015){Paudel}, {Duc}, \& {Ree}}]{Paudel15}
{Paudel}, S., {Duc}, P.~A., \& {Ree}, C.~H. 2015, \aj, 149, 114,
  \dodoi{10.1088/0004-6256/149/3/114}

\bibitem[{{Paudel} {et~al.}(2014){Paudel}, {Lisker}, {Hansson}, \&
  {Huxor}}]{Paudel14}
{Paudel}, S., {Lisker}, T., {Hansson}, K.~S.~A., \& {Huxor}, A.~P. 2014,
  \mnras, 443, 446, \dodoi{10.1093/mnras/stu1171}

\bibitem[{{Paudel} \& {Ree}(2014)}]{Paudel14b}
{Paudel}, S., \& {Ree}, C.~H. 2014, ArXiv e-prints.
\newblock \doarXiv{1410.7848}

\bibitem[{{Paudel} \& {Sengupta}(2017)}]{Paudel17b}
{Paudel}, S., \& {Sengupta}, C. 2017, \apjl, 849, L28,
  \dodoi{10.3847/2041-8213/aa95bf}

\bibitem[{{Paudel} {et~al.}(2013){Paudel}, {Duc}, {Cote}, {Cuillandre},
  {Ferrarese}, {Ferriere}, {Gwyn}, {Mihos}, {Vollmer}, {Balogh}, {Carlberg},
  {Boissier}, {Boselli}, {Durrell}, {Emsellem}, {MacArthur}, {Mei},
  {Michel-Dansac}, \& {van Driel}}]{Paudel13}
{Paudel}, S., {Duc}, P.-A., {Cote}, P., {et~al.} 2013, ArXiv e-prints.
\newblock \doarXiv{1302.6611}

\bibitem[{{Paudel} {et~al.}(2017){Paudel}, {Smith}, {Duc}, {C{\^o}t{\'e}},
  {Cuillandre}, {Ferrarese}, {Blakeslee}, {Boselli}, {Cantiello}, {Gwyn},
  {Guhathakurta}, {Mei}, {Mihos}, {Peng}, {Powalka}, {S{\'a}nchez-Janssen},
  {Toloba}, \& {Zhang}}]{Paudel17}
{Paudel}, S., {Smith}, R., {Duc}, P.-A., {et~al.} 2017, \apj, 834, 66,
  \dodoi{10.3847/1538-4357/834/1/66}

\bibitem[{{Pearson} {et~al.}(2016){Pearson}, {Besla}, {Putman}, {Lutz},
  {Fern{\'a}ndez}, {Stierwalt}, {Patton}, {Kim}, {Kallivayalil}, {Johnson}, \&
  {Sung}}]{Pearson16}
{Pearson}, S., {Besla}, G., {Putman}, M.~E., {et~al.} 2016, \mnras, 459, 1827,
  \dodoi{10.1093/mnras/stw757}

\bibitem[{{Privon} {et~al.}(2017){Privon}, {Stierwalt}, {Patton}, {Besla},
  {Pearson}, {Putman}, {Johnson}, {Kallivayalil}, {Liss}, \&
  {Titans}}]{Privon17}
{Privon}, G.~C., {Stierwalt}, S., {Patton}, D.~R., {et~al.} 2017, \apj, 846,
  74, \dodoi{10.3847/1538-4357/aa8560}

\bibitem[{{Putman} {et~al.}(2003){Putman}, {Staveley-Smith}, {Freeman},
  {Gibson}, \& {Barnes}}]{Putman03}
{Putman}, M.~E., {Staveley-Smith}, L., {Freeman}, K.~C., {Gibson}, B.~K., \&
  {Barnes}, D.~G. 2003, \apj, 586, 170, \dodoi{10.1086/344477}

\bibitem[{{Rhee} {et~al.}(2017){Rhee}, {Smith}, {Choi}, {Yi}, {Jaff{\'e}},
  {Candlish}, \& {S{\'a}nchez-J{\'a}nssen}}]{Rhee17}
{Rhee}, J., {Smith}, R., {Choi}, H., {et~al.} 2017, \apj, 843, 128,
  \dodoi{10.3847/1538-4357/aa6d6c}

\bibitem[{{Rich} {et~al.}(2012){Rich}, {Collins}, {Black}, {Longstaff}, {Koch},
  {Benson}, \& {Reitzel}}]{Rich12}
{Rich}, R.~M., {Collins}, M.~L.~M., {Black}, C.~M., {et~al.} 2012, \nat, 482,
  192, \dodoi{10.1038/nature10837}

\bibitem[{{Schlafly} \& {Finkbeiner}(2011)}]{Schlafly11}
{Schlafly}, E.~F., \& {Finkbeiner}, D.~P. 2011, \apj, 737, 103,
  \dodoi{10.1088/0004-637X/737/2/103}

\bibitem[{{Sengupta} {et~al.}(2014){Sengupta}, {Scott}, {Dwarakanath},
  {Saikia}, \& {Sohn}}]{Sengupta14}
{Sengupta}, C., {Scott}, T.~C., {Dwarakanath}, K.~S., {Saikia}, D.~J., \&
  {Sohn}, B.~W. 2014, \mnras, 444, 558, \dodoi{10.1093/mnras/stu1463}

\bibitem[{{Sengupta} {et~al.}(2012){Sengupta}, {Scott}, {Verdes Montenegro},
  {Bosma}, {Verley}, {Vilchez}, {Durbala}, {Fern{\'a}ndez Lorenzo}, {Espada},
  {Yun}, {Athanassoula}, {Sulentic}, \& {Portas}}]{Sengupta12}
{Sengupta}, C., {Scott}, T.~C., {Verdes Montenegro}, L., {et~al.} 2012, \aap,
  546, A95, \dodoi{10.1051/0004-6361/201219948}

\bibitem[{{Smith} {et~al.}(2007){Smith}, {Struck}, {Hancock}, {Appleton},
  {Charmandaris}, \& {Reach}}]{Smith07}
{Smith}, B.~J., {Struck}, C., {Hancock}, M., {et~al.} 2007, \aj, 133, 791,
  \dodoi{10.1086/510350}

\bibitem[{{Spergel} {et~al.}(2007){Spergel}, {Bean}, {Dor{\'e}}, {Nolta},
  {Bennett}, {Dunkley}, {Hinshaw}, {Jarosik}, {Komatsu}, {Page}, {Peiris},
  {Verde}, {Halpern}, {Hill}, {Kogut}, {Limon}, {Meyer}, {Odegard}, {Tucker},
  {Weiland}, {Wollack}, \& {Wright}}]{Spergel07}
{Spergel}, D.~N., {Bean}, R., {Dor{\'e}}, O., {et~al.} 2007, \apjs, 170, 377,
  \dodoi{10.1086/513700}

\bibitem[{{Springel} {et~al.}(2005){Springel}, {White}, {Jenkins}, {Frenk},
  {Yoshida}, {Gao}, {Navarro}, {Thacker}, {Croton}, {Helly}, {Peacock}, {Cole},
  {Thomas}, {Couchman}, {Evrard}, {Colberg}, \& {Pearce}}]{Springel05}
{Springel}, V., {White}, S.~D.~M., {Jenkins}, A., {et~al.} 2005, \nat, 435,
  629, \dodoi{10.1038/nature03597}

\bibitem[{{Stierwalt} {et~al.}(2015){Stierwalt}, {Besla}, {Patton}, {Johnson},
  {Kallivayalil}, {Putman}, {Privon}, \& {Ross}}]{Stierwalt15}
{Stierwalt}, S., {Besla}, G., {Patton}, D., {et~al.} 2015, \apj, 805, 2,
  \dodoi{10.1088/0004-637X/805/1/2}

\bibitem[{{Struck}(1999)}]{Struck99}
{Struck}, C. 1999, \physrep, 321, 1, \dodoi{10.1016/S0370-1573(99)00030-7}

\bibitem[{{Struck} \& {Smith}(2012)}]{Struck12}
{Struck}, C., \& {Smith}, B.~J. 2012, \mnras, 422, 2444,
  \dodoi{10.1111/j.1365-2966.2012.20798.x}

\bibitem[{{Tal} {et~al.}(2009){Tal}, {van Dokkum}, {Nelan}, \&
  {Bezanson}}]{Tal09}
{Tal}, T., {van Dokkum}, P.~G., {Nelan}, J., \& {Bezanson}, R. 2009, \aj, 138,
  1417, \dodoi{10.1088/0004-6256/138/5/1417}

\bibitem[{{Toloba} {et~al.}(2014){Toloba}, {Guhathakurta}, {van de Ven},
  {Boissier}, {Boselli}, {den Brok}, {Falc{\'o}n-Barroso}, {Hensler}, {Janz},
  {Laurikainen}, {Lisker}, {Paudel}, {Peletier}, {Ry{\'s}}, \&
  {Salo}}]{Toloba14}
{Toloba}, E., {Guhathakurta}, P., {van de Ven}, G., {et~al.} 2014, \apj, 783,
  120, \dodoi{10.1088/0004-637X/783/2/120}

\bibitem[{{Toomre} \& {Toomre}(1972)}]{Toomre72}
{Toomre}, A., \& {Toomre}, J. 1972, \apj, 178, 623, \dodoi{10.1086/151823}

\bibitem[{{van Dokkum}(2005)}]{vanDokkum05}
{van Dokkum}, P.~G. 2005, \aj, 130, 2647, \dodoi{10.1086/497593}

\bibitem[{{Yozin} \& {Bekki}(2012)}]{Yozin12}
{Yozin}, C., \& {Bekki}, K. 2012, \apjl, 756, L18,
  \dodoi{10.1088/2041-8205/756/1/L18}

\end{thebibliography}
\end{document}